\documentclass[12pt,a4paper,final]{iopart}

\usepackage{iopams}  
\usepackage{graphicx}
\usepackage[breaklinks=true,colorlinks=true,linkcolor=blue,urlcolor=blue,citecolor=blue]{hyperref}
\usepackage{subfigure}

\begin{document}

\title[Numerical evidences of universal trap-like aging dynamics]{Numerical evidences of universal trap-like aging dynamics}

\author[cor1]{Chiara Cammarota$^{1,2}$}
\ead{chiara.cammarota@kcl.ac.uk}

\author{Enzo Marinari$^2$}
\ead{enzo.marinari@uniroma1.it}

\vspace{0.8cm}
\address{$^1$Mathematics Department, King's College London, Strand London, United Kingdom}
\address{$^2$Dipartimento di Fisica, Universit\'a di Roma La Sapienza, 
CNR Nanotec unit\'a di Roma, INFN Sezione di Roma 1, Italy}

\begin{abstract}
Trap models have been initially proposed as toy models for dynamical relaxation in extremely simplified rough potential energy landscapes. Their importance has considerably grown recently thanks to the discovery that the trap like aging mechanism is directly controlling the out-of-equilibrium relaxation processes of more sophisticated spin models, that are considered as the solvable counterpart of real disordered systems. Establishing on a firmer ground the connection between these spin model out-of-equilibrium behavior and the trap like aging mechanism would shed new light on the properties, still largely mysterious, of the activated out-of-equilibrium dynamics of disordered systems. In this work we discuss numerical evidences of emergent trap-like aging behavior in a variety of disordered models. Our numerical results are backed by analytic derivations and heuristic discussions. Such exploration reveals some of the tricks needed to analyze the trap behavior in spite of the occurrence of secondary processes, of the existence of dynamical correlations and of finite system's size effects. 
\end{abstract}

\vspace{2pc}

\section{Introduction}
The study of extremely simplified spin models like the Random Energy Model (REM) and the $p$-spin model has been instrumental to the understanding of non-trivial statistical properties of super-cooled liquids and in general of complex systems in various fields \cite{cascav05}.  \\
As far as the dynamical behavior is concerned, the regime where complex systems always remain out-of-equilibrium and undergo aging is of paramount importance for practical applications.  Yet there is very little understanding about the underlying dynamical processes that determine the observable behavior.  For the spin models mentioned before a good understanding has been reached only in the case in which dynamical relaxation does not involve the crossing of energy barriers (non activated dynamics) \cite{bocukm98}.  For activated dynamics mean field computations remain of little use because in the large size limit, where saddle point solutions can be obtained, barriers diverge and cannot be crossed. Even for mean field spin models we are left with an essential lack of information on how the dynamics would look like on the time scales where barriers are crossed for finite system sizes.\\
The only way towards the description of activated dynamics has been to make a step backward and focus the attention on even simpler models. An example of paramount relevance has been the class of the \textit{trap models}.  In trap models the only element retained in the description of the slow dynamics is the time to escape from a deep minimum of the energy in the space of configurations, a.k.a. trap. Initially this mechanism was useful to provide an illustration of how aging could be originated \cite{boucha92}. Soon it became the new starting point to obtain quantitative predictions about aging dynamics for more sophisticated trap models \cite{rimabo00,rimabo01,montus03,beboga02} and qualitative descriptions of viscous liquids dynamics \cite{dyre1987,dyre1995}, until, very recently, the trap paradigm has been instrumental to achieve an unedited understanding of the REM dynamics \cite{cerwas15,gayrar16} and there are hopes that this insight might be extended to the dynamics of the $p$-spin model \cite{beboga02}.\\
An interesting open question is how much the trap aging paradigm is extended and whether it could become useful as a theoretical framework to describe out-of-equilibrium dynamics of super cooled liquids, glasses or other real systems for which it is believed that barrier crossing by thermal activation constitute the predominant dynamical mechanism \cite{cavagn09,berbir11}. Numerical simulations could be used to answer this question about systems for which it is hard to handle the analytic solution.  Nonetheless, as it emerges in the case of REM \cite{gayrar16} and as we are going to point out here, the underlying trap mechanism could remain hidden because decorated by secondary processes \cite{gayrar16,cammar15}, or washed out by dynamical correlations \cite{cammar15,babica17} due to small system sizes and finite observation times one is often bound to use in numerical simulations. In this note we study numerically the emergent trap dynamics features in two models (Poisson and Gaussian trap models) and we implement a few dynamical algorithms to reveal the difficulties one can meet in this task. To complement our many numerical results we will analyze some analytic predictions, through a review and some extensions of known results, and we will analyze in details how numerical data have to be analyzed and understood to reveal the underlying trap behavior will. Traps turn out to be, indeed, far more ubiquitous than one would have thought. \\
The structure of the paper is as follows: we recall the definition of trap models in \ref{Trap}, review the derivation of the long time aging dynamics features contained in the Arcsin law in \ref{predictionsTrap}, and comment about its numerical evidences in \ref{numericsTrap}.
We re-propose an extension of the trap dynamics where single transition rates depend on the energy of initial and final configurations of the dynamical step by rederiving the Arcsin law in \ref{predictionsaTrap} and discussing numerical evidences in \ref{numericsaTrap}.
We discuss how previous results change in the case of a Gaussian distribution of energies commenting about the predictions in \ref{predictionsGTrap} and \ref{predictionsaGTrap} and about their numerical tests in \ref{numericsGTrap} and \ref{numericsaGTrap}.
The last section is devoted to wrap up the study of generalised trap dynamics by comparing their outcomes with those of the Metropolis algorithm \ref{a0.5GTrapSM}. 

\section{The trap model}\label{Trap}
The dynamical process originally discussed by Bouchaud \cite{boucha92,boudea94} encodes in a schematic trap structure the effects of free-energy barriers and of the minima of a rough potential energy landscape. Traps are represented by single configurations sitting at the bottom of deep holes and surrounded by energy barriers that hamper quick relaxation in every direction. 
Each instance $i$ of the $M$ allowed 
configurations is associated to a energy $E_i<0$ (that does not change in time). 
Its surrounding barriers always reach a fixed high threshold energy $E_b$. 
Before changing configuration, the dynamics is held in trap $i$ for a time lapse given by a Poisson random variable with average 
\begin{equation}
\tau_i=\tau_0\exp[\beta (E_b-E_i)]
\label{timeBT}
\end{equation}
(the average Arrhenius time for a stochastic process with white Gaussian noise of variance $2/\beta$ to climb a barrier $E_b-E_i$). 
The next trap $j$ is then chosen uniformly among all the available  traps: 
$p_{i,j}=1/M$.
In summary the transition rates between configurations $i$ and $j$ are 
\begin{equation}
r_{i,j}=\frac{1}{M}\exp[-\beta (E_b-E_i)]\ .
\label{rateBT}
\end{equation} 
In the trap model this dynamics is applied \cite{boucha92,boudea94} to a system where 
the barrier height is set to zero: $E_b=0$ and the energies are obtained from $M$ independent realizations $\epsilon_i$ of a Poisson random variable $\epsilon$ with rate $\lambda$:
\begin{equation*}
\phi_P(\epsilon;\lambda)=\lambda\exp(-\lambda\epsilon)
  \mbox{ and }E_i = -\epsilon_i\;.
\end{equation*}
The assumption of a Poisson distribution of the energies, rather than the more natural Gaussian distribution (that characterizes other simple and relevant spin glass models like REM \cite{derrid81}), was meant to retain only the effect of the deepest configurations.
The minima of large collections of Gaussian energies are indeed Gumbel distributed, so that the left tail vanishes exponentially (i.e. like the right tail of the Poisson distribution). \\
The crucial consequence of the fact that trap models are described by the absolute values of the energies distributed with Poisson is that the mean trapping times are distributed with a fat tailed-power law:  
\begin{equation}
\rho(\tau;\lambda/\beta)=\frac{\lambda}{\beta}\;\frac{\tau_0}{\tau^{1+\frac{\lambda}{\beta}}} \ .
\label{rhotau}
\end{equation}
When $\lambda<\beta$ the average trapping time is infinite and the long time behavior of the system  is dominated by the trapping time in the deepest traps found. This inherently excludes the possibility of equilibration (large time limits for one-time observable quantities cannot be defined) and implies an aging dynamics \cite{boucha92,boviha94,boudea94}.\\
The out-of-equilibrium dynamics arising in Bouchaud trap model can be understood analytically in very good detail. Interestingly, this dynamics can be seen as a starting point towards the study of a single spin flip dynamics or molecular glass dynamics. However, trap models simplify a realistic aging dynamics in many ways.\\
First, in a trap model energies of neighboring configurations (or traps) are independent from each other. This is justified by the idea that different configurations in trap models are not simple neighboring configurations but schematically represent different minima separated by barriers in the potential energy landscape. Still it is not evident even that energies of neighboring minima are completely uncorrelated.\\
Second, once a trap is left any other configuration can be reached in a single step. In other words there is not a realistic structure of dynamically connected configurations (this is what would happen in a dynamics where one can flip any number of spins at a time, as opposed to a single spin flip dynamics). Moreover the abundance of escape directions hampers the possibility of recurrent visits to the same trap, bringing to zero the correlation between subsequent configurations visited by the dynamics.\\
Third, the dynamical evolution of the system in every possible direction requires to go across potential energy barriers of the same height: this simplifies the actual complexity of the energy landscape. \\
Finally at each step the energy always goes back to the threshold energy $E_b$ adding to the usual Markov dynamical process the important property of being {\it renewal}. In fact since this threshold does not depend on the trap that the system just left, and the new trap is uniformly chosen among all the traps, every step is totally independent from the previous one and can be treated as a new start for the whole dynamics. The resulting dynamics is at the same time invariant under translation and inversion symmetry, nonetheless the system ages.

\subsection{An insightful derivation of the Arcsin law}
\label{predictionsTrap}
During aging the limit $\lim_{t_w\rightarrow\infty} C(t_w,t_w+t)$ of the two time correlation function $C(t_w,t_w+t)$ is a non-trivial function $\mathcal{C}(\omega)$ of the variable $\omega=t/t_w$ \cite{bocukm98}. The calculation that we discuss here (following \cite{boudea94}) focuses on the probability $\Pi[t_w,t_w+t]$ that the dynamical process does not jump in a time interval between $t_w$ and $t$\footnote{There are other two point functions that show aging behavior. In general the presence or absence of aging and its features depends on the exact function one uses. The function $\Pi$ is able to distinguish true aging from the sub-aging behavior that occurs in $1d$ models~\cite{berbou03,bencer05}.}. We will show here that
\begin{equation}
\lim_{t_w\rightarrow\infty,\ t/t_w=\omega} \Pi[t_w,t_w+t] = H_{\lambda/\beta}(\omega) \;,
\end{equation}
where
\begin{equation}
H_{x}(\omega)\equiv\frac{\sin(\pi x)}{\pi}\int_{\omega}^{\infty} \frac{du}{(1+u)\;u^x} \;,
\end{equation}
and $x=\lambda/\beta$.\\
For a system in configuration $i$ with trapping time $\tau_i$, the probability of not jumping away before time $t$ is 
\begin{equation}
\Pr[\Theta>t;\tau_i] \equiv \int_t^{\infty} \frac{dt'}{\tau_i} \exp(-t'/\tau_i) = \exp(-t/\tau_i) \ ,
\end{equation}
where $\Theta$ is a random variable representing the escape time with average $\tau_i$ specified in Eq. (\ref{timeBT}). The total probability $\Pi[t_w,t_w+t]$ of not having a jump between $t_w$ and $t_w+t$ is given by sum over all the configurations $i$ of the probability that at time $t_w$ the system is in the configuration $i$ (with trapping time $\tau_i$), $\mathrm{p}[\tau_i;t_w]$, times the probability $\Pr[\Theta>t;\tau_i]$ that the system will not jump within $t$:
\begin{equation}
\Pi[t_w,t_w+t]=\sum_{i=1}^M \mathrm{p}[\tau_i;t_w] \Pr[\Theta>t;\tau_i] \ . 
\end{equation} 
The probability of having a trapping time $\tau_i$ obeys the recursive relation: 
\begin{eqnarray}
\mathrm{p}[\tau_i;t_w]&=& \frac{1}{M}\int_{t_w}^{\infty}\frac{dt}{\tau_i} \exp(-t/\tau_i)\nonumber\\ 
&+& \frac{1}{M}\sum_{j=0}^M \int_0^{t_w}\frac{dt}{\tau_j} \exp(-t/\tau_j) \mathrm{p}[\tau_i;t_w-t] \ ,
\label{Pitw}
\end{eqnarray}
i.e. the system is in trap $i$ at time $t_w$ either because it never jumped away before $t_w$ or because it was in any configuration $j$ (included $i$) up to a time $t<t_w$, it jumped away at time $t$ and it is found to be in trap $i$, with trapping time $\tau_i$, after a time $t_w-t$. Note here the importance of the renewal property of the dynamics: $p[\tau_i;t-t_w]$ is the temporal shift of the left hand side. If the dynamics is completely independent from previous steps a temporal shift does not affect this term. Hence, under this assumption it is possible to find an explicit expression for the probability $p[\tau_i;t]$. We only need to transform the convolution in a product by using the Laplace transform.  \\
The Laplace transform of $\Pi[t_w,t_w+t]$ is 
\begin{eqnarray}
\hat{\Pi}(s,t) &\equiv& \int_0^\infty dt_w \exp(-st_w)\; \Pi[t_w,t_w+t] \nonumber \\
&=& \sum_{j=0}^M \Pr[\Theta>t;\tau_i] \int_0^\infty dt_w \exp(-st_w)\mathrm{p}[\tau_i;t_w] \ .
\end{eqnarray}
Using Eq. (\ref{Pitw}) we find
\begin{eqnarray}
\hspace{-2.5cm}
\int_0^\infty dt_w \exp(-st_w)\;\mathrm{p}[\tau_i;t_w]&=&\int_0^\infty dt_w \exp(-st_w)\frac{1}{M}\int_{t_w}^{\infty}\frac{dt}{\tau_i} \exp(-t/\tau_i) \\
&+& \int_0^\infty dt_w \exp(-st_w) \frac{1}{M}\sum_{j=0}^M \int_0^{t_w}\frac{dt}{\tau_j} \exp(-t/\tau_j) \mathrm{p}[\tau_i;t_w-t] \ ,\nonumber 
\end{eqnarray}
where the first term on the right hand side gives 
\begin{equation}
\int_0^\infty dt_w \exp(-st_w)\frac{1}{M}\int_{t_w}^{\infty}\frac{dt}{\tau_i} \exp(-t/\tau_i)=\frac{1}{M}\frac{\tau_i}{1+s\tau_i}\;.
\end{equation}
By performing a change of variables and multiplying and dividing by $\exp(-st)$ the second term becomes
\begin{eqnarray}
\hspace{-2.5cm}
&\frac{1}{M}&\int_0^\infty dt_w \exp(-st_w) \sum_{j=0}^M \int_0^{t_w}\frac{dt}{\tau_j} \exp(-t/\tau_j) \mathrm{p}[\tau_i;t_w-t]= \\ \nonumber
&\frac{1}{M}&\sum_{j=0}^M \int_0^{\infty} \frac{dt}{\tau_j} \exp(-t/\tau_j)\exp(-st)\int_0^\infty d(t_w-t)\exp(-s(t_w-t))\mathrm{p}[\tau_i;t_w-t] = \\ \nonumber
&\frac{1}{M}&\sum_{j=0}^M\frac{1}{1+s\tau_j}\int_0^\infty dt_w \exp(-st_w)\mathrm{p}[\tau_i;t_w] \ .
\end{eqnarray}
Putting together these results we get
\begin{equation}
\hspace{-2.5cm}
\int_0^\infty dt_w \exp(-st_w)\mathrm{p}[\tau_i;t_w] =\frac{1}{M}\frac{\tau_i}{1+s\tau_i}+\frac{1}{M}\sum_{j=0}^M\frac{1}{1+s\tau_j}\int_0^\infty dt_w \exp(-st_w)\mathrm{p}[\tau_i;t_w]
\end{equation}
and hence 
\begin{equation}
\int_0^\infty dt_w \exp(-st_w)\mathrm{p}[\tau_i;t_w] =\frac{\frac{\tau_i}{1+s\tau_i}}{s\sum_{j=0}^M\frac{\tau_j}{1+s\tau_j}} \ .
\end{equation}
Since by definition $\rho(\tau;x)\equiv\sum_{j=0}^M \delta(\tau-\tau_j)/M$, in the limit of large $M$ we can approximate the sum over configurations with an integral on the trapping times weighted by their distribution and we get
\begin{equation}
s\sum_{j=0}^M\frac{\tau_j}{1+s\tau_j}=\int^{\infty}_{0}\sum_{j=0}^M\delta(\tau-\tau_j)\frac{s\tau}{1+s\tau}\simeq\int^{\infty}_{0}d\tau M\rho(\tau;x)\frac{s\tau}{1+s\tau} \ .
\end{equation}
By substituting the expression of the trapping time distribution (that we know for $\tau>\tau_0$) we obtain, for very small $\tau_0$ and setting $x=\lambda/\beta$,
\begin{equation}
\hspace{-1.5cm}
s\sum_{j=0}^M\frac{\tau_j}{1+s\tau_j}\simeq \int^{\infty}_{\tau_0}d\tau M\frac{x\tau_0^x}{\tau^{1+x}}\frac{s\tau}{1+s\tau}\simeq Mx\tau_0^xs^x\Gamma(x)\Gamma(1-x)=\frac{\pi M x \tau_0^xs^x}{\sin(\pi x)}  \ ,
\end{equation}
that allows to derive the final expression for the Laplace transform of $\mathrm{p}[\tau_i;t_w]$:
\begin{equation}
\int_0^\infty dt_w \exp(-st_w)\mathrm{p}[\tau_i;t_w] =\frac{\sin(\pi x)}{\pi M x \tau_0^x s^x}\frac{\tau_i}{1+s\tau_i} \ .
\end{equation}
We can now evaluate the Laplace transform of the correlation function. We approximating again the sum over configurations with the integral over trapping times
\begin{equation}
\hat{\Pi}(s,t) = \int_{\tau_0}^\infty d\tau \frac{Mx\tau_0^x}{\tau^{1+x}} \int_t^{\infty}\frac{dt'}{\tau}\exp(-t/\tau') \frac{\sin(\pi x)}{\pi M x \tau_0^x s^x} \frac{\tau}{1+s\tau} \ ,
\end{equation}
which after a number of simplifications, changes of variables ($u=s\tau$ and $t_w=t'/u$) and integration ranges, becomes, for small $\tau_0$ 
\begin{equation}
\hat{\Pi}(s,t) \simeq \int_{0}^\infty d t_w \exp(-st_w) \frac{\sin(\pi x)}{\pi}\int_{t/t_w}^{\infty} du \frac{1}{u^x(1+u)}  \ .
\end{equation}
By simply comparing the last equation with the definition of $\hat{\Pi}(s,t)$ as the  Laplace transform of $\Pi[t_w,t_w+t]$ we finally get
\begin{equation}
\Pi[t_w,t_w+t] \simeq \frac{\sin(\pi x)}{\pi}\int_{t/t_w}^{\infty} du \frac{1}{u^x(1+u)}  \ ,
\end{equation}
with $x=\lambda/\beta$.
The last result goes under the name of Arcsin law\footnote{From a technical point of view, the Arcsin law is originated by the large time convergence of the dynamical process to a stable subordinator (see for example \cite{beboga02,bencer06,beboce08,pityor92} and references therein).}.

\subsection{Numerical observation of the Arcsin law} \label{numericsTrap}
\begin{figure}[htb!]
 \centering   
 \subfigure[\label{AG_EXP_ADS_T0.50_a0.00_w0.10} ]{\includegraphics[width=0.32\columnwidth, angle=-90]{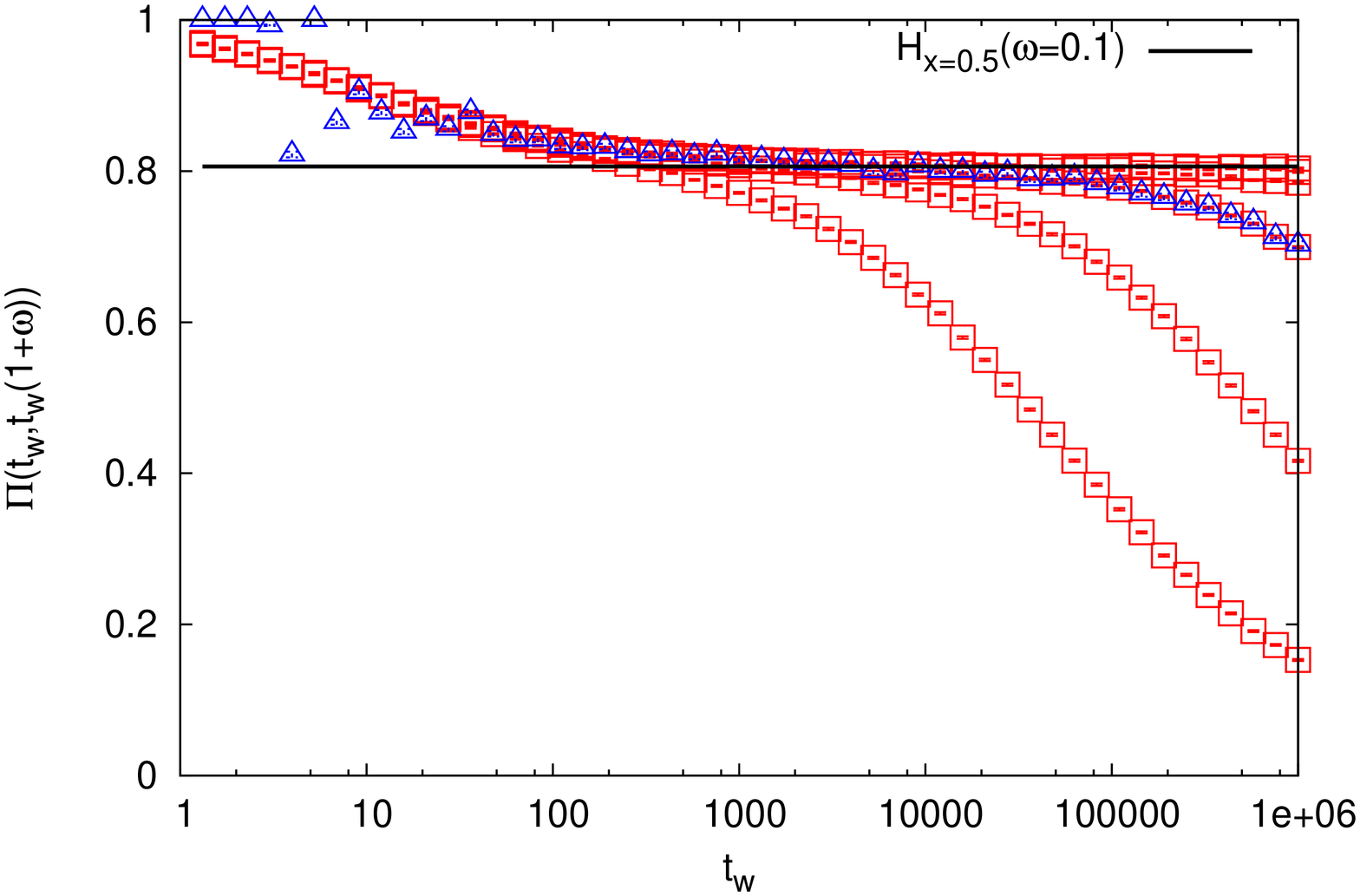}}
 \subfigure[\label{AG_EXP_ADS_T0.50_a0.00_w2.50} ]{\includegraphics[width=0.32\columnwidth, angle=-90]{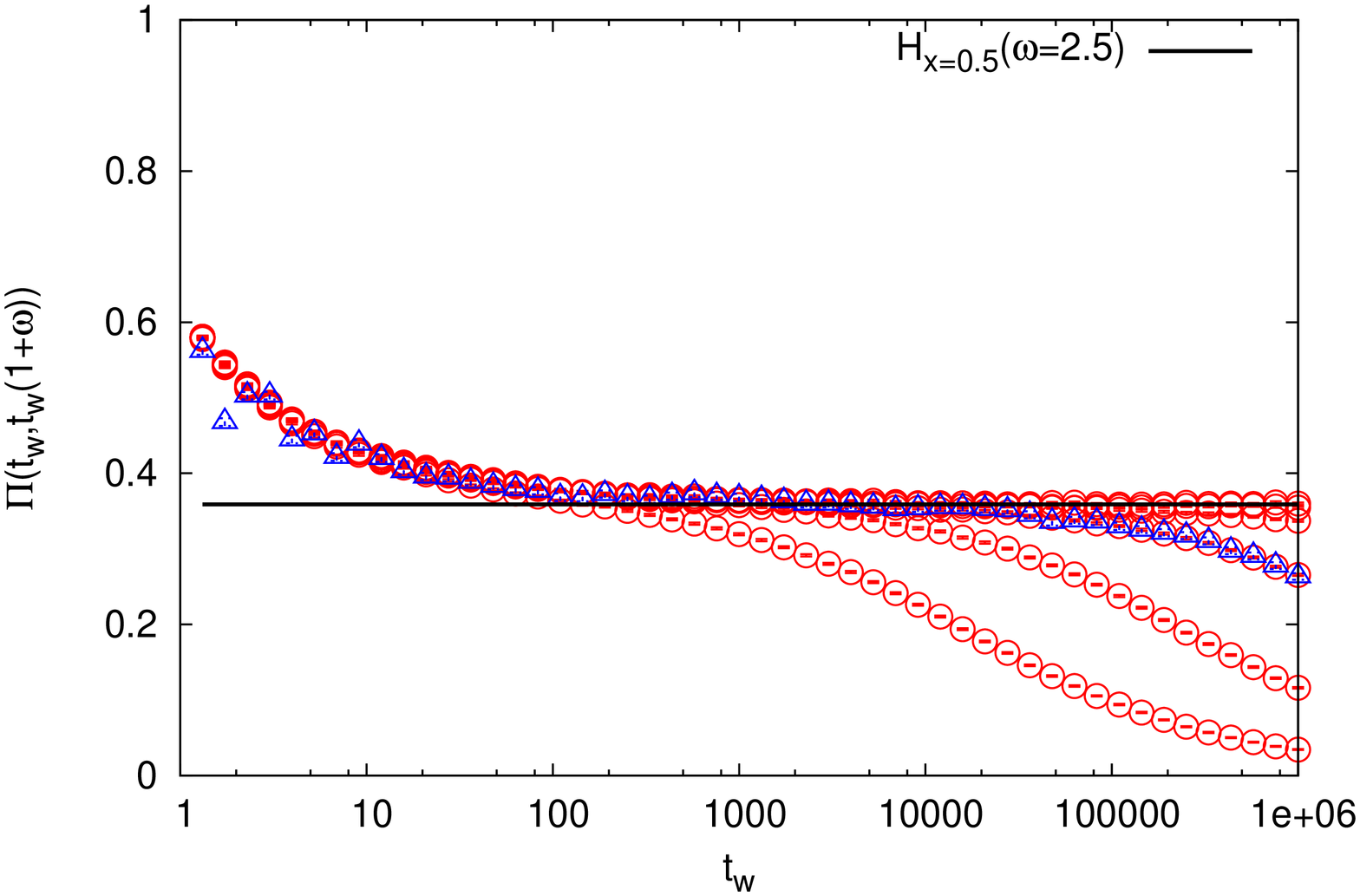}}
 \caption{$\Pi_M[t_w,t_w(1+\omega)]$ with $\omega=0.1$ in \ref{AG_EXP_ADS_T0.50_a0.00_w0.10} and $\omega=2.5$ in \ref{AG_EXP_ADS_T0.50_a0.00_w2.50} for a CTD at $T=0.5$. Different data series, plotted with red circles, are for CTD, with, from left to right, $M=64$, $256$, $1024$, $4096$, $16384$, $65536$.\label{AG_EXP_ADS_T0.50_a0.00} Triangles are for DTD, $M=1024$, $10000$ samples, and for large $t_w$ they overlap the $M=1024$ CTD data. The straight black lines are for the theoretical predictions.}
\end{figure}

We have run numerical simulations of both a continuous time dynamics (CTD) based on an average jumping time from trap $i$, defined in (\ref{timeBT}), and a discrete time dynamics (DTD) based on transition rates between trap $i$ and trap $j$, defined in (\ref{rateBT}). The two dynamics allow exploring the rough potential energy landscape of a trap model as defined at the beginning of this section with $E_b=0$ and a Poisson distribution of energy absolute values with rate $\lambda=1$.\\
In the CTD, at each step we increment the time of the dynamics by a Poisson random variable with average given in Eq. (\ref{timeBT}), and we extract uniformly the new configuration.  In the DTD, at each step of the dynamics we increment the time of the process by $\Delta t$ and we extract a target configuration uniformly among all the available ones. We extract a number $r$ according to a uniform distribution between $0$ and $1$ and we compare it with a normalized transition rate $r^N_{i,j} = r_{i,j} / (\max_{i,j} r_{i,j})$, where $r_{i,j}$ is given by (\ref{rateBT}): if $r<r^N_{i,j}$ the move is accepted, otherwise it is rejected and the proposed target configuration is disregarded (a new target configuration will be chosen for the next step).  The
elementary time scale $\tau_0$ of the CTD and the elementary time step $\Delta t$ of the DTD are set to $1$. The dynamics is repeated for different realizations of the potential energy landscape ($10^5$ times for CTD and $10^4$ times for DTD)\footnote{In general to allow for a comparison of the two  dynamics (CTD and DTD) we will need to properly set the time scale  in accordance with the convention adopted for the normalization of the transition rates. In the present case the rescaling leads to  unit factors. We will come back to this point in the next sections  when the introduction of non trivial factors will be needed.}.\\
We have used CTD to study systems of different sizes, with $M=64$, $256$,
$1024$, $4096$, $16384$, $65536$ at a temperature $T=0.5$. We have
measured the probability $\Pi[t_w,t_w+t]$ of not having a jump between $t_w$ and $t_w+t$ on logarithmic intervals of $t_w$ and values of $t/t_w=\omega=0.1$, $2.5$.  We plot the results as a function of $t_w$ for different system sizes.\\
Fig. \ref{AG_EXP_ADS_T0.50_a0.00_w0.10} shows $\Pi_M[t_w,t_w(1+\omega)]$ with $\omega=0.1$ for CTD with $T=0.5$.
Fig. \ref{AG_EXP_ADS_T0.50_a0.00_w2.50} is for CTD at $T=0.5$  with  $\omega=2.5$.
The probability of not jumping in the time interval shows a short time transient and equilibration at large $t_w$ for small enough system sizes. 
When the system size increases equilibration occurs at larger time scales and a plateau develops. This shows that in the large system size and large $t_w$ limit the probability of not jumping $\Pi[t_w,t_w(1+\omega)]$ is accurately estimated by $H_{\lambda/\beta}(\omega)$.
Both figures show a good agreement between the fitted plateau levels for the largest system size and the analytic prediction.\\
The DTD gives compatible results. In this case the numerical simulation times are larger than for CTD, so we only report results about one system size $M=1024$ averaged over $10^4$ samples for the two values of $\omega$ in Fig. \ref{AG_EXP_ADS_T0.50_a0.00}.\\
The conclusion of this first exploratory study is that simulations of the trap model show very clearly (already for very small systems like the one with $M=256$ configurations) the emergent plateau of $\Pi[t_w,t_w+t]$ which is in good agreement with the trap aging predictions. We will see that this is not always the case for other dynamics, even if they are also expected to be governed by a trap like mechanism. In those cases a more careful analysis of the numerical results will be needed to let the trap aging behavior to emerge.

\section{The \texorpdfstring{$a$}{a}-generalized trap model}\label{aTrap}
The results that we have discussed in the previous section can be extended in many different ways to explain the phenomenology of more realistic models of glass dynamics.  First heuristically~\cite{monbou96} and eventually rigorously~\cite{beboga02,beboga03,beboga03bis} different authors have analyzed the emergence of a trap-like dynamics in models where the energies of configurations are Gaussian distributed (we will come back to this topic in the next section) or even correlated with each other~\cite{beboga02,beboce08,bedegu01}, mimicking the typical energy distribution of hard spin configurations in $p$-spin models with $p\ge3$.\\
However, as we have already discussed, other simplifications characterize the trap paradigm. For example the dynamics is such that the depth of the traps is only determined by the energy level of the initial configuration. As a consequence the barrier to be crossed by the system in order to change configuration does not depend on the chosen direction or dynamical path. \\
As we said before, this feature is directly related with the renewal property of the aging dynamics. Yet, it is strongly at variance with usual dynamical processes (the Metropolis algorithm being a typical and effective example) or with what we expect from a realistic description of glass dynamics, where different microscopic moves imply different energetic costs.\\
In this section we focus on a generalization of the continuous time dynamics of trap models. It aims  at gradually including a dependence from the transition rate on the energy of the target configuration, as it has already been done in some previous works  \cite{rimabo00,rimabo01,montus03}. We will discuss in detail the theoretical setting, numerically check and analyze it and we will extend it to different situations.\\
A generalized trap dynamics \cite{rimabo00,rimabo01,montus03} deviates from the classical one due to the introduction of an interpolating parameter $a\in [0,1]$. The transition rates from an initial configuration $i$ to a final configuration $j$ read
\begin{equation}
r_{i,j}\propto \exp[\beta(1-a)E_i-\beta a E_j] \ .
\end{equation}
When $a=0$ one recovers the original trap model. When $a=1/2$, the transition rates become very similar to the ones of a Metropolis dynamics (in the sense that the two energies have the same influence on the dynamical step) with a temperature that is double of the one that would govern Metropolis). The prefactor $1-a$ to the $E_i$ term is essential for the detailed balance to hold. \\
The same dynamics can be mapped onto a continuous time dynamics that generalizes the usual trap paradigm by assuming that the time spent in a configuration $i$ followed by the final configuration $j$ is given by a trapping time scale 
\begin{equation}
\tau_i=\tau_0\exp[\beta(E_b-(1-a)E_i)]
\label{timeaBT}
\end{equation}
(the actual trapping time being a Poisson random variable with average $\tau_i$) and introducing a non uniform probability to get to the final configuration $j$
\begin{equation}
p_{i,j}=\frac{1}{Z^{(a)}}\exp(-\beta a E_j) 
\label{pija}
\end{equation}
with $Z^{(a)}=\sum_j\exp(-\beta a E_j)$.
In particular this set up corresponds to the previous discrete time dynamics with 
\begin{equation}
r_{i,j}=\frac{1}{Z^{(a)}} \exp[-\beta(E_b-(1-a)E_i+a E_j)] \ .
\label{rija}
\end{equation}
Note that once the dynamics has reached a configuration $i$, this configuration will trap on average the dynamics during a time $\tau_i$.
However thanks to the uneven sampling of the final configurations, the same configuration $i$ will be sampled along the dynamics with a frequency 
$p_{j,i}\propto\exp(-\beta a E_i)$, enforcing for  the configuration $i$ the usual Boltzmann weight: 
$\tau_i\exp(-\beta a E_i)\propto\exp(-\beta E_i)$. 
The numerical implementation of the two dynamics is usually faster after a rescaling of the transition rates $r_{i,j}$ and $p_{i,j}$ such that their maximum is set to one. When $a\neq 0$ this maximum will depend on the sample, and in the case of Gaussian  distribution of energies also on the system size. To allow the comparison between the two dynamics and, in the next sections, to avoid a spurious scaling of the resulting time scale of the dynamics we need to rescale by the same factor the unit time of the DTD and $\tau_0$ in the CDT.\\
For the time being, like for the classical trap dynamics studied so far, we will still consider a system with negative energies with absolute values sampled from a Poisson distribution with rate $\lambda$ and a reference level for the barriers heights $E_b=0$.

\subsection{Derivation of the Arcsin law for an $a$-generalized trap dynamics}
\label{predictionsaTrap}
Following the derivation of the Arcsin law for the original trap dynamics it is easy to show \cite{rimabo01}, through a derivation of the trapping time distribution, that a similar trap-like behavior describes the aging process of an $a$-generalized trap model, provided the parameter $x$ of the Arcsin law is correctly evaluated as a combination of $\lambda$, $T$, and the interpolating parameter $a$. \\
Since the absolute values of the energies are Poisson distributed with rate $\lambda$ and the average trapping times are proportional to $\exp(-\beta(1-a)E)$, instead of $\exp(-\beta E)$, the trapping time distribution is $\rho\left(\tau;\frac{\lambda}{\beta(1-a)}\right)$.
Moreover due to the non uniform sampling of the energies along the dynamics, according to $p_{i,j}$ in Eq. (\ref{pija}), the Laplace transform of the probability of having trapping time $\tau_i$ at time $t_w$, $\mathrm{p}[\tau_i;t_w]$, will become 
\begin{equation}
\int_0^\infty dt_w \exp(-st_w)\mathrm{p}[\tau_i;t_w] =\frac{\left(\frac{\tau_i}{\tau_0}\right)^{\frac{a}{1-a}}\frac{\tau_i}{1+s\tau_i}}{s\sum_{j=0}^M\left(\frac{\tau_j}{\tau_0}\right)^{\frac{a}{1-a}}\frac{\tau_j}{1+s\tau_j}} 
\end{equation}
with denominator 
\begin{equation}
s\sum_{j=0}^M\left(\frac{\tau_j}{\tau_0}\right)^{\frac{a}{1-a}}\frac{\tau_j}{1+s\tau_j}\simeq M\frac{\pi x' \tau_0^xs^x}{\sin(\pi x)}
\end{equation}
where $x'=\frac{\lambda}{\beta(1-a)}$ and $x=\frac{(\lambda/\beta-a)}{(1-a)}=x'-\frac{a}{(1-a)}$. Hence
\begin{equation}
\int_0^\infty dt_w \exp(-st_w)\mathrm{p}[\tau_i;t_w] =\frac{\sin(\pi x)}{M\pi x' \tau_0^{x'} s^x}\tau_i^{\frac{a}{1-a}}\frac{\tau_i}{1+s\tau_i} \ ,
\end{equation}
\begin{equation}
\hat{\Pi}(s,t) = \int_{\tau_0}^\infty d\tau \frac{x'\tau_0^{x'}}{\tau^{1+x'}} \int_t^{\infty}\frac{dt'}{\tau}\exp(-t/\tau')\frac{\sin(\pi x)}{\pi x' \tau_0^{x'} s^x}\tau^{\frac{a}{1-a}}\frac{\tau}{1+s\tau} \ ,
\end{equation}
and
\begin{equation}
\Pi[t_w,t_w+t] \simeq \frac{\sin(\pi x)}{\pi}\int_{t/t_w}^{\infty} du \frac{1}{u^x(1+u)}  \ ,
\end{equation}
with $x=(\lambda/\beta-a)/(1-a)$. \\
A final comment on this result is that the $a$ generalization of the parameter of the Arcsin law can be immediately interpreted if we think that the uneven sampling of the energies along the dynamics acts as if the distribution of the energy traps had a lower rate $\lambda'=\lambda-\beta a$ compared to the original one. Besides, the prefactor $1-a$ at the exponent of the trapping times can be included in the definition of a rescaled temperature $\beta'=\beta(1-a)$.
The $a$-generalized dynamics can hence be mapped into a classical trap dynamics with rescaled rate $\lambda'$ of the Poisson energies and rescaled temperature $1/\beta'$, whose aging is described by an Arcsin law with parameter $$x=\lambda'/\beta'=(\lambda/\beta-a)/(1-a) \ .$$

\subsection{Numerical checks of the  \texorpdfstring{$a$}{a}-generalized trap-like dynamics}
\label{numericsaTrap}
\begin{figure}[htb!]
 \centering   
 \subfigure[\label{AG_EXP_ADS_T0.50_a0.25_w0.10} ]{\includegraphics[width=0.32\columnwidth, angle=-90]{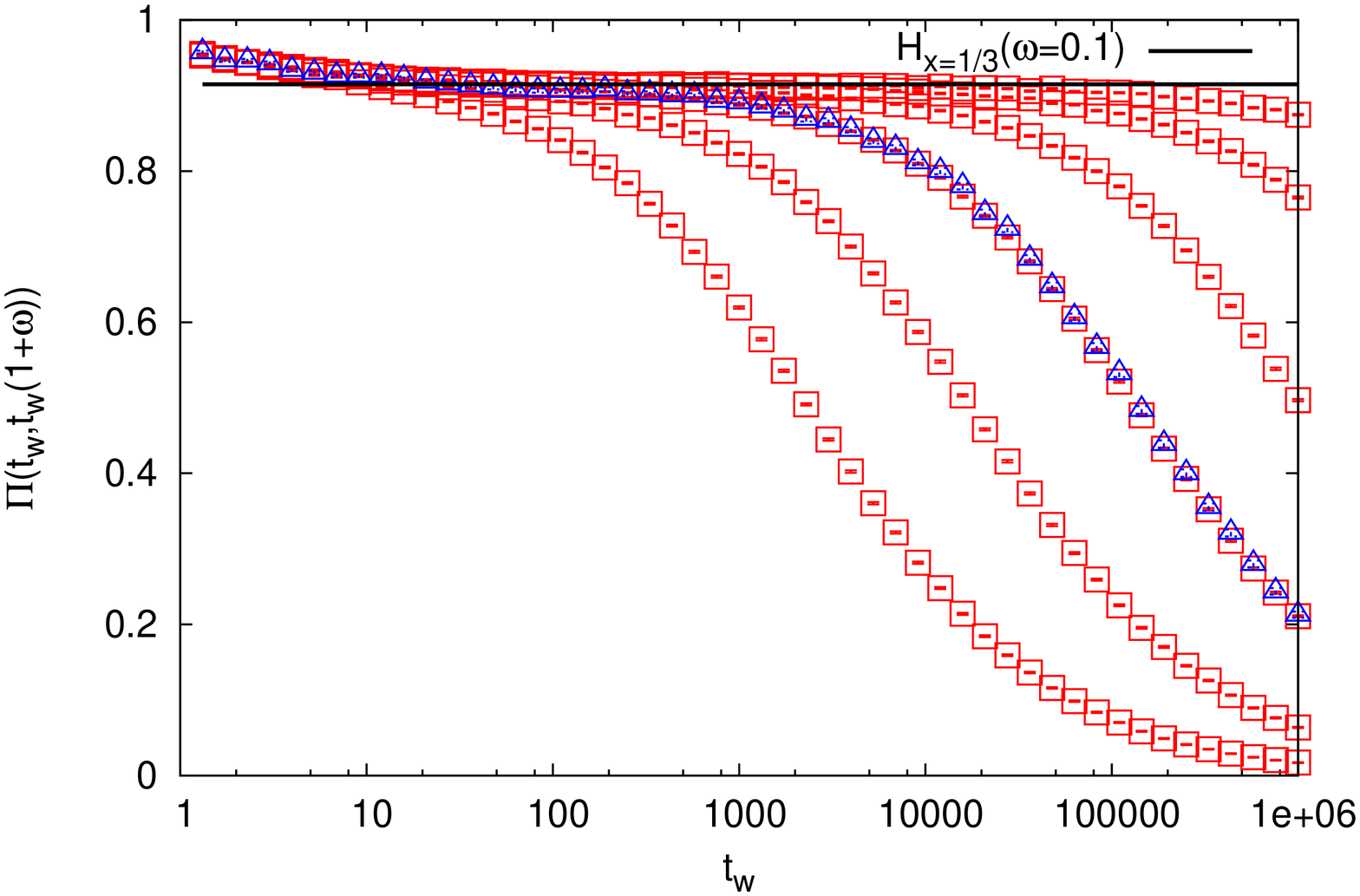}}
 \subfigure[\label{AG_EXP_ADS_T0.50_a0.25_w2.50} ]{\includegraphics[width=0.32\columnwidth, angle=-90]{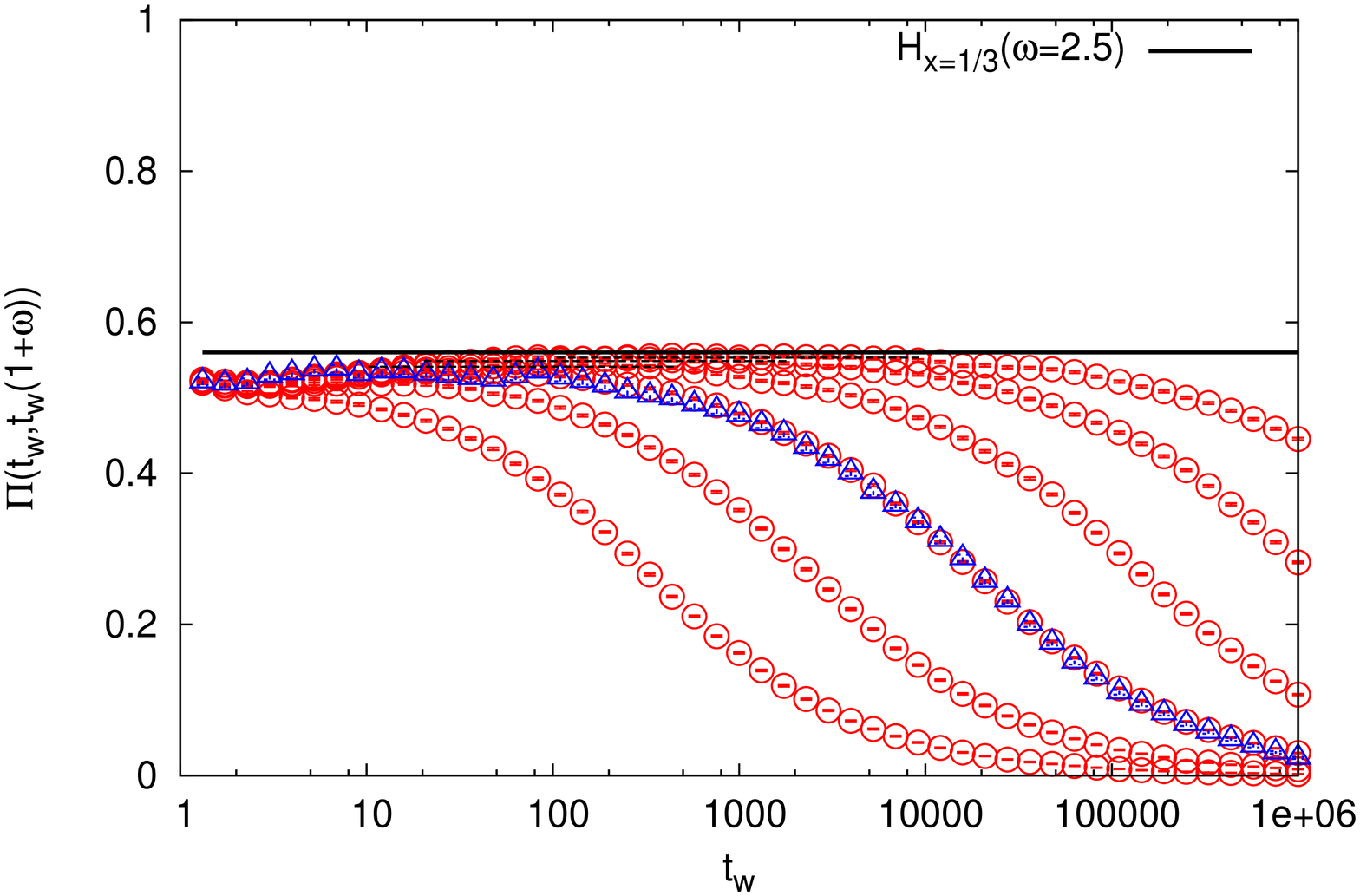}}
 \caption{$\Pi_M[t_w,t_w(1+\omega)]$ with $\omega=0.1$ in \ref{AG_EXP_ADS_T0.50_a0.25_w0.10} and $\omega=2.5$ in \ref{AG_EXP_ADS_T0.50_a0.25_w2.50} for an $a$-generalized CTD with $a=0.25$ at $T=0.5$. Different data series correspond to different system sizes: from left to right $M=64$, 256, 1024, 4096, 16384, 65536.\label{AG_EXP_ADS_T0.50_a0.25}  
 Triangles are for DTD, $M=1024$, 10000 samples, and for large $t_w$ they overlap the $M=1024$ CTD data.}
\end{figure}

We have run numerical simulations of the $a$-generalized trap dynamics by using the $a$-generalized CTD and DTD. In the CTD the trapping time in Eq. (\ref{timeaBT}) sets the timescale for the elementary time of single steps while the system resides in the configuration $i$. Attempts to change configuration are accepted with the rate in Eq. (\ref{pija}) that depends on the target configuration only. In the DTD attempts to change configurations are accepted with a rate given in Eq. (\ref{rija}) and $\Delta t=1$.
The results for $T=0.5$, $a=0.25$, and different sizes $M=64$, 256, 1024, 4096, 16384 and 65536 have been averaged over $100000$ samples. They are shown as a function of $t_w$ in Fig.\ref{AG_EXP_ADS_T0.50_a0.25_w0.10} for $\omega=0.1$ and in Fig.\ref{AG_EXP_ADS_T0.50_a0.25_w2.50} for $\omega=2.5$.
The correlation function of  Fig.\ref{AG_EXP_ADS_T0.50_a0.25_w2.50} is in some cases non-monotonic. We can see that a flat region emerges  even for small system sizes (see the levels of the short dashed segments in the figure), and it is lower than the asymptotic plateau predicted by the theory. In fact the trap aging prediction is reached by the plateau only in the limit of very large system sizes. A very careful finite size scaling study must be used not only to detect the appearance of a plateau but also to show its convergence to the level it would actually have in the large system size limit. Comparison with trap expectations is appropriate only in this large time asymptotic region.\\
We also simulated the $a$-generalized DTD. This again requires a larger simulation time. The results coincide with the ones obtained with CTD.  We show in Fig. \ref{AG_EXP_ADS_T0.50_a0.25} the results for  $M=1024$ averaged  over $10000$ samples.

\section{The Gaussian trap model}\label{GTrap}
A natural choice for the energies distribution of systems with $N$ interacting degrees of freedom, for large $N$, is the Gaussian distribution with $\sigma=N$
\begin{equation}
\phi_N(E;N) = \frac{1}{\sqrt{2\pi N}}\exp\left(-\frac{E^2}{2N}\right)\ .
\end{equation}
In this case energies can be seen as the result of the sum of a large number of independent microscopic contributions. Moreover a Gaussian distribution of energies is the main feature of the simplest model that is believed to retain essential thermodynamic features of the glassy behavior, the random energy model (REM) \cite{derrid81}. From a thermodynamic perspective this model shows a transition from a paramagnetic high-temperature phase to a glassy low-temperature phase where the partition function is dominated by the configuration associated with the smallest energy. The transition temperature is $T_c=1/\beta_c=1/\sqrt{2\log 2}$. \\
While the statics of this simple model has been wholly understood, reaching a deep and complete understanding of its dynamical behavior is not easy. Very recent rigorous results show that the Arcsin law governs the long time out of equilibrium behavior of the REM \cite{cerwas15,gayrar16}: this implies that REM belongs to the trap like class since aging dynamics. However, from a numerical point of view it is not straightforward to obtain evidence of this expected trap-like aging behavior \cite{babica17}: we will analyze here this issue. Before focusing on the numerical results we need to specify the kind of dynamics we will be working with and the theoretical expectations on the aging behavior implied by this dynamics.\\
The dynamics of the REM model, considered as an extreme simplification of a spin model, should be characterized by a non trivial structure of dynamically connected configurations.  Single spin flip dynamics allows any starting configuration $i$ to be surrounded only by $N$ neighbors. In other words only $N$ configurations can be reached in one single dynamical step. More formally, the dynamics can be represented as the walk of a point on the $2^N$ vertices of an hyper-cube in $N$ dimensions with $2$ vertices per side, where every single vertex is always connected to $N$ other vertices.\\
To keep our new model closer to the original trap model, we consider instead trap models where the distribution of energies is Gaussian, but a single dynamical move can connect any initial configuration $i$ with every other one \cite{beboga02}. We will call these models Gaussian trap models (GTM). Again in the simplest version of GTM single configurations are assumed to be surrounded by barriers of height $E_b=0$ so that the transition rates governing the dynamics are the ones defined in Eq. (\ref{rateBT}), or equivalently the dynamics can be described by a continuous time process with trapping times given by Eq. (\ref{timeBT}). As we will see the trap dynamics of a system with Gaussian energies produces qualitatively different $\Pi[t_w,t_w+t]$ compared with Poisson trap models. Furthermore their behavior will be very similar with the one obtained in studies of the actual REM dynamics \cite{babica17}. As such, even within the simplified settings of a GTM, we will obtain, and discuss how to interpret, non trivial numerical results.

\subsection{The Arcsin law emerges on properly rescaled time scales}\label{predictionsGTrap}
It has been shown~\cite{beboga02} that for a continuous time process on the GTM with trapping times described by Eq. (\ref{timeBT}), the
probability $\Pi[t_w, t_w+t]$ of staying in the same configuration between $t_w$ and $t_w + t$ shows the
same aging behavior as the one of Bouchaud trap model, in the limit of large system's size and after a specific rescaling of the observation times.
In fact it has been shown that 
\begin{equation}
\lim_{t_w\rightarrow \infty,t/t_w=\omega} \lim_{N\rightarrow \infty}\Pi[\theta(N) t_w, \theta(N) (t_w+t)]=H_x(\omega)
\end{equation}
where $\theta(N)=\exp(\gamma N)$ is the exponential rescaling factor, $x=\sqrt{2\rho\log 2}/\beta$, and $\gamma=\beta^2 x$. \\
As it will become clearer in the following, the observation time must be rescaled to recover the standard trap-like aging dynamics because even if we are considering Gaussian distributed energies, the minima of large collections of these energies will be characterized by a distribution with Poisson tails. Note that this was one of the original motivations for the choice of a Poisson distribution of energies in the Bouchaud trap models. \\
Let us discuss an intuitive explanation of the need of observing the dynamics at rescaled times and a prediction about the aging behavior that should emerge at the rescaled times. We can consider the dynamics in the Gaussian trap model as a sequence of explorations of groups of $m=2^{\rho N}$ configurations (with $\rho\in(0,1)$). As long as $\beta>\sqrt{\rho}\beta_c$ the configuration with the smallest energy within each of these groups will be trapping the dynamics for most of the time spent in the set. This is due to the fact that for $\beta>\sqrt{\rho}\beta_c$ the sum of the trapping times will be dominated by the largest element: $\sum_{i=1}^m\tau_i\sim 2^{N\rho}\int dE\exp(-E^2/2-\beta E)$. Its saddle point solution is $E_{SP}=-N\beta$ for $\beta<\sqrt \rho \beta_c$, and gets stuck at $\overline E_{\rm min}=-N\sqrt \rho \beta_c$ for $\beta>\sqrt \rho \beta_c$. Hence in the second regime $\sum_{i=1}^m\tau_i\sim\max_{i}\tau_i=\tau(\overline E_{\rm min})$ with $\overline E_{\rm min}=\overline {\min_i E_i}$. Moreover the distribution of the minima encountered in each of these groups of $m$ configurations is Poissonian \cite{boumez97}. In general \cite{boumez97} for a Gaussian distribution of the original energies $E_i$ with mean $\mu$ and variance $N$ the typical minimum energy of one of $m$ variables is $\overline E_{\rm min}=\mu-\sqrt{2N}\sqrt{\log m}$ and the deviations from this average are described by random variables distributed according to a Gumbel distribution, with a Poisson left tail, with rate $\lambda=\sqrt{2\log m/N}$. \\
In the present case $\mu=0$ and $m=2^{\rho N}$ hence $\overline E_{\rm min}=-N\sqrt{\rho}\sqrt{2\log 2}$ and $\lambda=\sqrt{\rho}\sqrt{2\log 2}$. This information allows a straightforward mapping of the dynamics of the Gaussian trap model onto the trap-like aging dynamics of the original Bouchaud trap model with a suitable choice of $\lambda$. On one side, the unit time of the effective trap dynamics must be the typical time needed to explore the bunch of $m$ configurations. This will be dominated by the time needed to escape from the lowest configuration: $\tau(\overline E_{\rm min})\sim\tau_0\exp(\beta \sqrt{\rho} \sqrt{2\log 2} N)$. To observe an effective trap like process we have to scale the observation time by this same factor $\tau(\overline E_{\rm min})\sim\theta(N)=\exp(\gamma N)$ with $\gamma=\beta\sqrt{\rho}\sqrt{2\log 2}$. Note that this time scale grows exponentially with the system size. On the other side, the configurations trapping the dynamics on this time scale will be Poisson distributed with $\lambda=\sqrt{\rho}\sqrt{2\log 2}$. Hence the Arcsin law will be describing the aging dynamics of the Gaussian trap model in the same way as for a Bouchaud trap models with parameter $x=\lambda/\beta=\sqrt{\rho}\sqrt{2\log 2}/\beta$, in agreement with the rigorous results reported at the beginning of this section. This intuitive explanation \cite{beboga02} of the mechanism for the emergence of a trap-like behavior in GTM is useful to interpret the numerical results. Moreover it is simple enough to be easily extended to the $a$-generalized dynamics and to provide intuition on the expected trap dynamics also in that case.\\
An important remark is that this results holds for any choice of $\rho\in(0,1)$. The larger is the value of $\rho$ we select, the larger is the timescale over which the dynamics must be observed. Similarly, when the aging process takes place, the longer we wait before observing its behavior the larger $\rho$ and the aging parameter $x$ will be, until in certain cases $x\rightarrow1$ and aging will stop.\\ 
In fact in GTM aging will be always interrupted at sufficiently large time scales according to two different mechanisms. Since $x=\sqrt{\rho}\sqrt{2\log 2}/\beta$,  the system is effectively behaving as a trap model with fixed $\lambda_{\rm eff}=\sqrt{2\log 2}=\beta_c$ and an effective temperature $T_{\rm eff}=T\sqrt{\rho}$, which is increasing  with time. In particular since $T_{\rm eff}=T\sqrt{\rho}$ for short enough times (small $\rho$) aging should be visible at any temperature, even for $T>T_c$ ($\beta<\beta_c$), as long as $T_{\rm eff}<T_c$. In this high temperature regime $\beta_{\rm eff}$ will decrease, with increasing time and $\rho$, and reach $\beta_c$ at $\rho_{\rm eq}=\beta/\beta_c<1$. At this point aging will stop because $x$ has reached the value one. The interruption of aging in this case takes place before the system has explored all the configurations at disposal ($\rho_{\rm eq}<1$ and $2^{\rho_{\rm eq}N}<M$). In fact the system equilibrates as soon as it has explored configurations at the equilibrium energy $-N\beta$. This happens at a time scale $\exp(\beta^2 N)$, exponentially diverging with the system size.\\
For $\beta>\beta_c$ instead, $\beta_{\rm eff}$ will decrease until $\rho\rightarrow 1$ where $\beta_{\rm eff}=\beta>\beta_c$ and hence still $x<1$. In this case aging stops because the system has sampled all the configurations at disposal including the ground state. This occurs at a timescale exponentially diverging with the system's size $\exp(\beta \sqrt{2\log 2}N)$ and depending on the ground state energy $-\sqrt{2\log 2}N$.

\subsection{Numerical evidence for trap-like aging in models with Gaussian energies}\label{numericsGTrap}
In this section we discuss the numerical results obtained with the CTD and the  DTD for systems where energies are Gaussian distributed. As in previous cases we set $E_b=0$. The time scale $\tau_0$ in Eq. (\ref{timeBT}) and the elementary time step $\Delta t$ for the DTD are such that the two dynamics can be compared even if transition rates are rescaled to optimize simulation times. This also assures that the dynamics in not altered by the introduction of spurious time scales growing exponentially with the system size.\\
The system sizes we consider are $N=6$, 8, 10, 12, 14 and 16, corresponding respectively to $M=64$, 256, 1024, 4096, 16384 and 65536 configurations. 
\begin{figure}[htb!]
 \centering   
 \subfigure[\label{AG_GAU_ADS_T1.58_a0.00_w0.10} ]{\includegraphics[width=0.32\columnwidth, angle=-90]{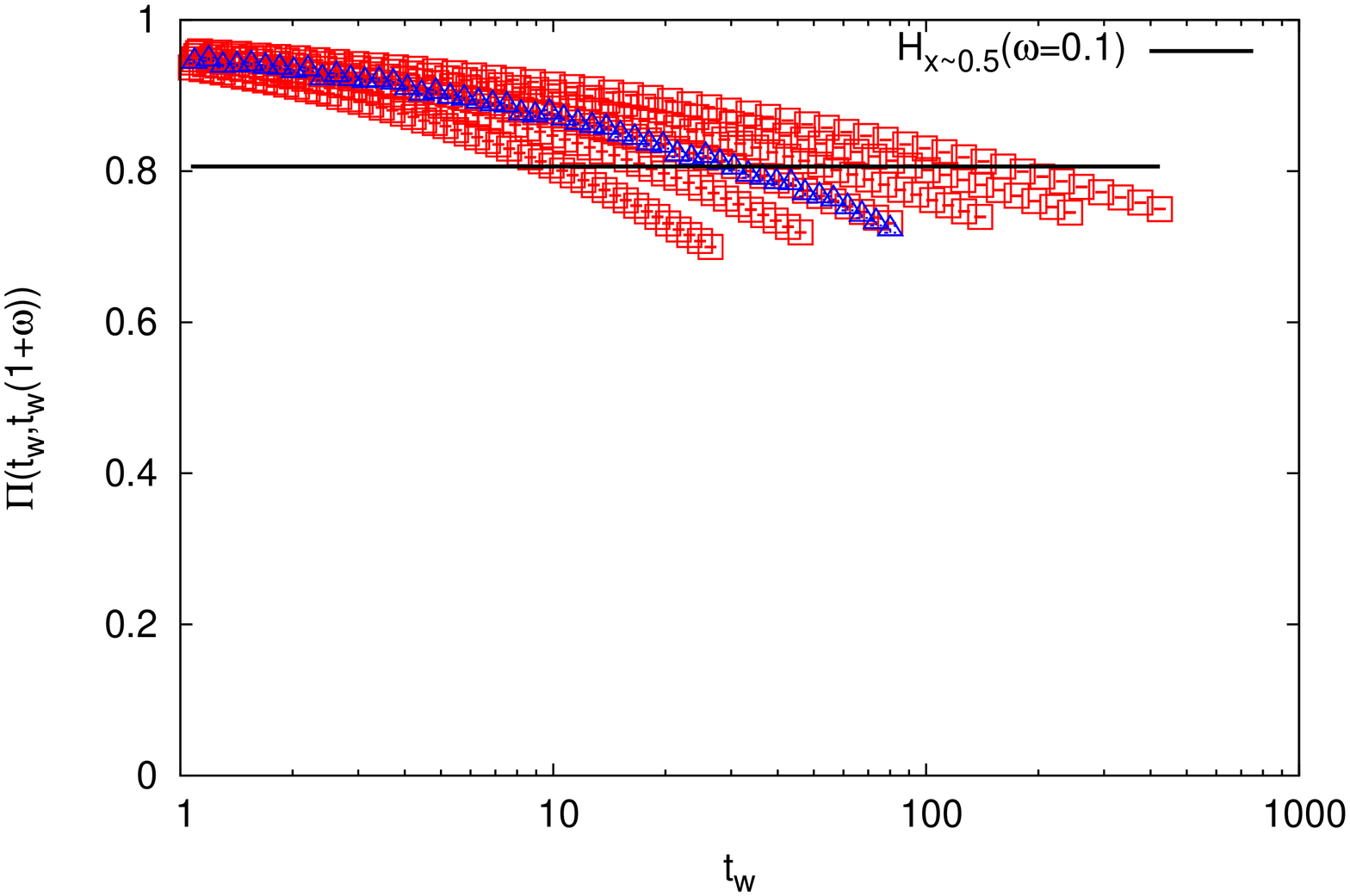}}
 \subfigure[\label{AG_GAU_ADS_T1.58_a0.00_w2.50} ]{\includegraphics[width=0.32\columnwidth, angle=-90]{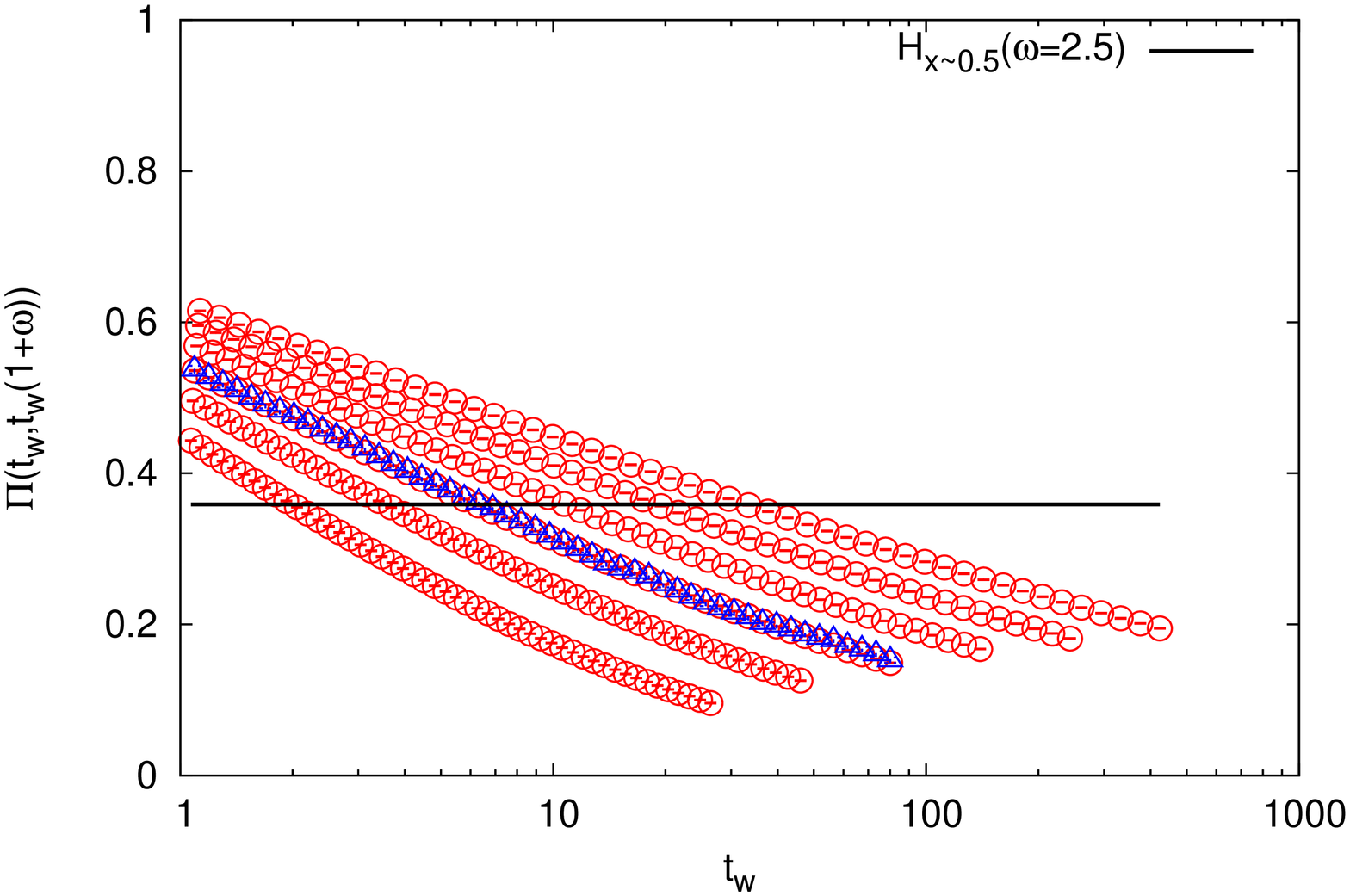}}
 \caption{$\Pi_M[t_w,t_w(1+\omega)]$ with $\omega=0.1$ in \ref{AG_GAU_ADS_T1.58_a0.00_w0.10} and $\omega=2.5$ in \ref{AG_GAU_ADS_T1.58_a0.00_w2.50} for a CTD at $T=1.58/\beta_c$ and $\rho=0.1$.  The chosen parameters correspond to $x\sim 0.5$. Different data series from left to right correspond to system sizes $N=6$, 8, 10, 12, 14 and 16, with $10^{6}$ samples.\label{AG_GAU_ADS_T1.58_a0.00} Superimposed triangles represent the same data obtained with DTD, $N=10$, averaged over $10^4$ samples, and for large $t_w$ they overlap the $N=10$ CTD data.}
\end{figure}
The time scale corresponds to the time needed to explore $M^{\rho}$ with $\rho=0.1$. We have used $T=1.58/\beta_c$ to get a trap like aging behavior with parameter $x=0.5(=\sqrt{\rho}\beta_c/\beta)$ for the probability of not having a jump between $t_w$ and $t_w+t$, $\Pi[t_w,t_w+t]$. We have averaged the CTD over $10^6$ samples. The results for $\omega=0.1$ and $\omega=2.5$ are shown in Fig.~\ref{AG_GAU_ADS_T1.58_a0.00}.\\
At variance with the case of Poisson distribution of energies we do not see the formation of a plateau. When time is increasing the system is going across infinitely many different aging regimes characterized by different values of the effective parameter $x$, as the temperature that would be appropriate for and effective trap description is continuously increasing. In this sense the resulting data describe a convolution of infinitely many plateau. The only change in the curve we can detect is that, for the largest system sizes, when data are shown in $\log(t_w)$ scale they show a decreasing slope. Even the curves with the smallest slope are still far from clearly revealing whether the aging dynamics is approaching the trap prediction (the black line in the figure) or not.\\
The perspective changes completely when we use the other piece of information we got from our phenomenological study and we plot data as function of time in units of the correct observation time scale $\exp(\gamma N)$.
\begin{figure}[htb!]
 \centering   
\subfigure[\label{AG_GAU_ADS_T1.58_a0.00_w0.10SC} ]{\includegraphics[width=0.32\columnwidth, angle=-90]{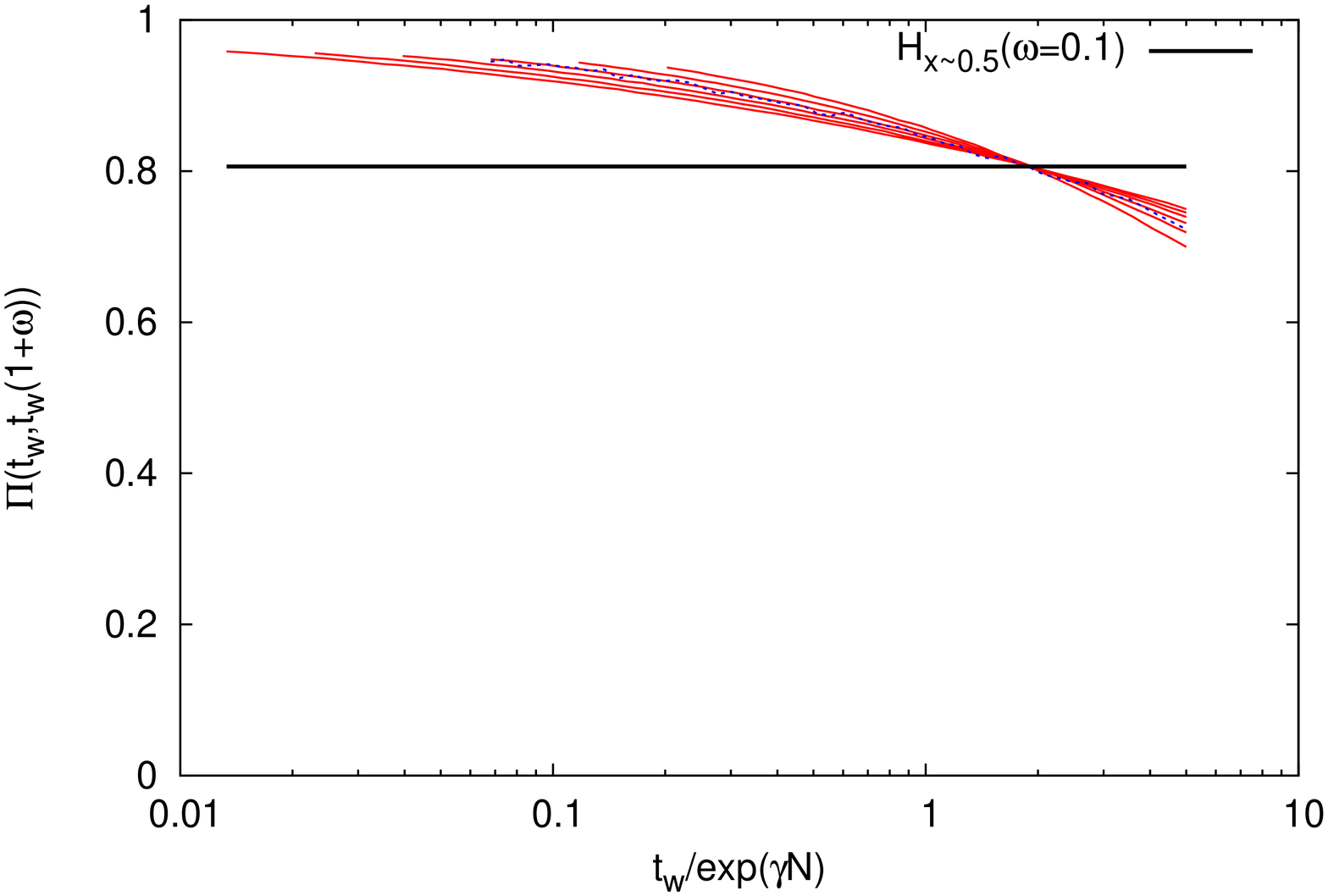}}
 \subfigure[\label{AG_GAU_ADS_T1.58_a0.00_w2.50SC} ]{\includegraphics[width=0.32\columnwidth, angle=-90]{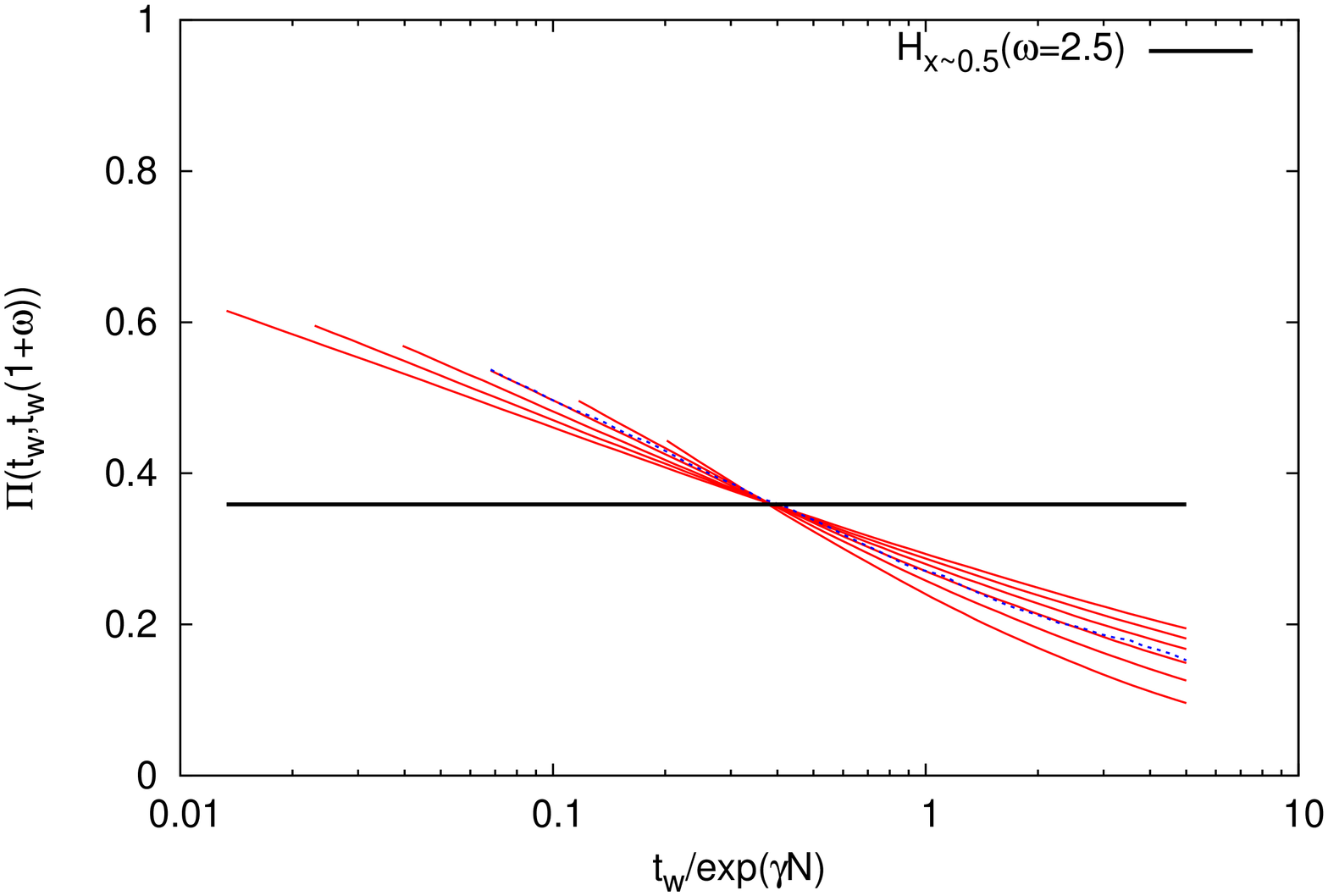}}
 \caption{The $\Pi_M[t_w,t_w(1+\omega)]$ data from the CTD, also shown in Fig. \ref{AG_GAU_ADS_T1.58_a0.00}, here as a function of $t_w/\exp(\gamma N)$. Small system sizes are on the right at the top and on the left at the bottom, and the opposite holds for large system sizes. The dashed line is for the results obtained with the DTD. \label{AG_GAU_ADS_T1.58_a0.00SC}}
\end{figure}
The result of this rescaling is in Fig.~\ref{AG_GAU_ADS_T1.58_a0.00SC}.
We use lines joining the data points 
to allow to distinguish between different data series.
The time rescaling clearly reveals where the large system size limit is placed: it is at the level where data series cross, that coincides with excellent precision with the theoretical prediction. The probability of not jumping 
decreases with $N$ above the crossing and increases below the crossing.  
The smaller slope for large system sizes is telling that the probability of not changing configuration is remaining unchanged for longer and longer times for increasing $N$.
We would need huge systems to determine the correct infinite size limit without assuming the correct scaling.
By looking at the dynamics of the right correct scales it becomes possible to get information about the asymptotic trap like aging behavior even from small systems.

\section{Gaussian trap model aging dynamics for an \texorpdfstring{$a$}{a}-generalized trap dynamics}\label{aGTrap}
We will now consider a further generalization of the standard trap dynamics, and we will introduce a dependence of the transition rates both on the initial and final configuration, similarly to what happens to standard Metropolis dynamics (this is the dynamical process that we have defined and discussed in Sec.~\ref{aTrap}). We discuss here the predictions we can make for the aging behavior of such a dynamics on the Gaussian trap model.\\
\subsection{The Arcsin law emerges on  rescaled time scales in the $a$-generalised dynamics}\label{predictionsaGTrap}
As usual the transition rate can be split in a trapping time scale as in Eq.~(\ref{timeaBT}), depending on the initial configuration, and a probability to get to the final configuration (Eq.~(\ref{pija})). Also in this case to get any hope to recover the trap like aging dynamics we need to be able to explore a number of configurations exponentially large in the system's size $N$. This allows  to exploit the extreme value statistics property of Gaussian energies and obtain an effective Poisson distribution of the energy of deep configurations. In particular the weight for final configurations, according to which the energy landscape is probed, allows an exponentially large number of equilibrium configurations only if $\beta a<\sqrt{2\log 2}$. This can be obtained by solving a REM with inverse temperature $\beta a$ and hence with the Boltzmann weight of the sampling probability (\ref{pija}). In this model the number of configuration that are typically explored at equilibrium is 
\begin{equation}
M(a)=\exp\left[N\left(\log 2-\frac{a^2\beta^2}{2}\right)\right] \ .
\end{equation}
This number is much smaller than $M=\exp(N\log 2)$ but it is still exponentially growing with the system size $N$ when $\beta a<\sqrt{2\log 2}$. Their mean energy is $\mu(a)=-N\beta a$. Their energy values are Gaussian distributed around $\mu(a)$ with variance $N$. Despite minor quantitative changes (concerning the mean and the number of available configurations) we are in the same qualitative situation as for the dynamics with $a=0$. We can hence consider the case of a dynamics observed on time scales that allow for the observation of a subgroup of $m(a)=M(a)^{\rho}$ configurations. The deepest configurations in different groups of $m(a)$ configurations will have average energy of $\overline E_{\rm min}=-N\beta a-\sqrt{\rho}\sqrt{2\log 2 - a^2\beta^2}$ and the left tail of their distribution will be Poisson with rate $\lambda=\sqrt{\rho}\sqrt{2\log 2-a^2\beta^2}$. Note that in the case $a=0$ the original $\overline E_{\rm min}$ and $\lambda$ are recovered, and the condition on the temperature becomes $\beta>\sqrt{\rho}\beta_c$ as previously stated.\\
The time spent in each single group of $m(a)$ configurations $\sum_{i=1}^{m(a)}\tau_i$ will be dominated by the time spent in the deepest configuration among them $\sum_{i=1}^{m(a)} \tau_i \sim \max_i \tau_i$ when 
\begin{equation}
\beta>\sqrt{\frac{2\log 2}{\left[\frac{(1-a)^2}{\rho}+a^2\right]}}=\frac{\beta_c}{\sqrt{\left[\frac{(1-a)^2}{\rho}+a^2\right]}} \ .
\label{bca}
\end{equation}
In fact at this point the saddle point energy of
\begin{equation*}
\sum_{i=1}^{m(a)}\tau_i\sim\int dE\exp\left[\rho N(\log 2-\frac{a^2\beta^2}{2})-\beta(1-a)E-\frac{(E+N\beta a)^2}{2N}\right]\;, 
\end{equation*}
$E_{SP}=-N\beta$, becomes equal to $\overline E_{\rm min}$, and the sum is dominated by its largest term, with minimum energy.\\
Using these results we can conclude that trap like aging dynamics is expected to emerge when $\beta<\sqrt{2\log 2}/a$ and under the condition in Eq. (\ref{bca}) on timescales proportional to $\exp(-\beta (1-a) \overline E_{\rm min})=\exp(\gamma N)$ with $\gamma=(1-a)\beta(a\beta+\sqrt{\rho}\sqrt{2\log 2-a^2\beta^2})$. The trap dynamics emerging on these time scales with effective inverse temperature $\beta(1-a)$ and exploring configurations with energies that effectively Poisson-distributed with rate $\lambda = \sqrt{\rho} \sqrt{2\log 2-a^2\beta^2}$ will be described by an Arcsin law with parameter 
\begin{equation}
    x=\frac{\lambda}{ \beta(1-a)}=\frac{\sqrt{\rho}\sqrt{2\log 2-a^2\beta^2}}{\beta(1-a)} \ .
\end{equation}
As for the simple trap model dynamics for the GTM, the longer is the waiting time of the system, the larger is the corresponding value of $\rho$. Eventually the system equilibrates due to the fact that either the dynamics had the time to explore all the available configurations (including the one with the lowest energy), or it has reached the typical equilibrium configuration at that temperature.\\
More precisely, the trap predictions for the $a$-generalized dynamics of GTM imply two limits for the range of validity of this result: $\beta=\sqrt{2\log 2}/a$ and the saturation of the condition in Eq.~(\ref{bca}). The first limit corresponds to getting $x=0$ for any $\rho$, hence any time scale. Aging will not be visible at low temperatures such that $\beta$ exceeds $\sqrt{2\log 2}/a$ because the system will be stuck in a sub-exponential number of configurations. For higher temperatures instead aging will be visible for any $a\in(0,1)$, at least for a finite range of time. In particular at the very beginning of the dynamics we will always start with $x=0$. On longer time scales, for fixed $a$, again we distinguish two ranges in temperature: if $\beta>\beta_c/\sqrt{(1-a)^2+a^2}$ (intermediate temperature) aging will continue till $x=\sqrt{2\log 2 -a^2 \beta^2}/\beta(1-a)$ and $\rho=1$ which occurs at the time when all the $M(a)$ available configurations have been explored and equilibration takes place. This time scale is $\exp[(1-a)\beta(a\beta+\sqrt{2\log 2-a^2\beta^2})N]$, and it is again exponentially growing with the system size. If instead $\beta<\beta_c/\sqrt{(1-a)^2+a^2}$ (high temperature) aging will be interrupted when $x$ becomes one at $\rho_{\rm eq} = (1-a)^2 \beta^2/ (\beta_c^2 -a^2 \beta^2)$ and at timescale $\exp[\beta^2(1-a)N]$, when the exploration of equilibrium configurations at energy $-\beta N$ has been completed.

\subsection{Numerical check of the trap-like dynamics for generic values \texorpdfstring{$a$}{a}}\label{numericsaGTrap}
We have analyzed numerically the $a$-generalized CDT and DTD.
Again, as in the $a=0$ case, we have studied the dynamics where $E_b$ is set to $0$ and microscopic time scales are set in such a way that it appears evident that CDT and DTD are equivalent.
\begin{figure}[htb!]
 \centering   
 \subfigure[\label{AG_GAU_ADS_T1.58_a0.25_w0.10} ]{\includegraphics[width=0.32\columnwidth, angle=-90]{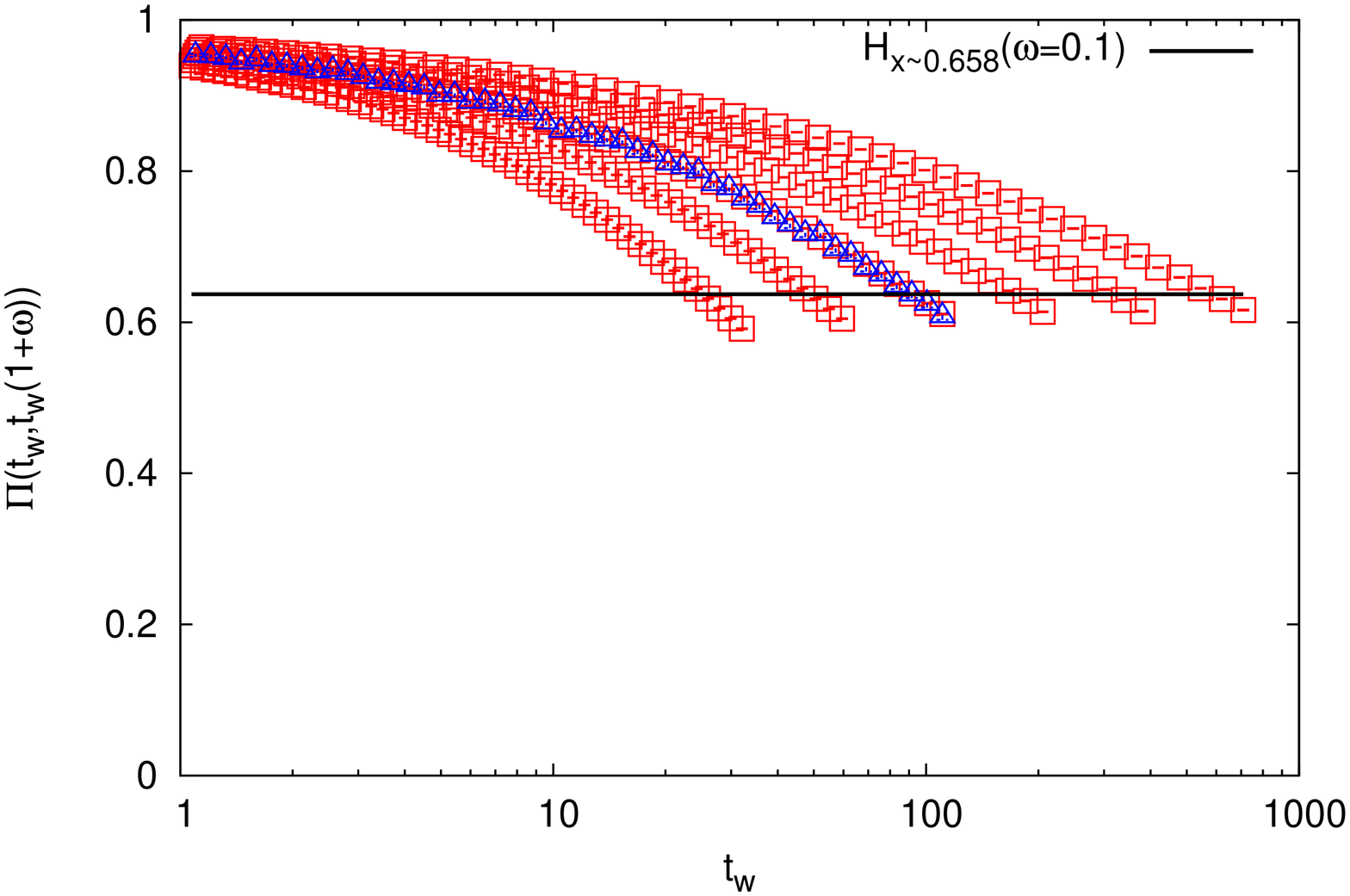}}
 \subfigure[\label{AG_GAU_ADS_T1.58_a0.25_w2.50} ]{\includegraphics[width=0.32\columnwidth, angle=-90]{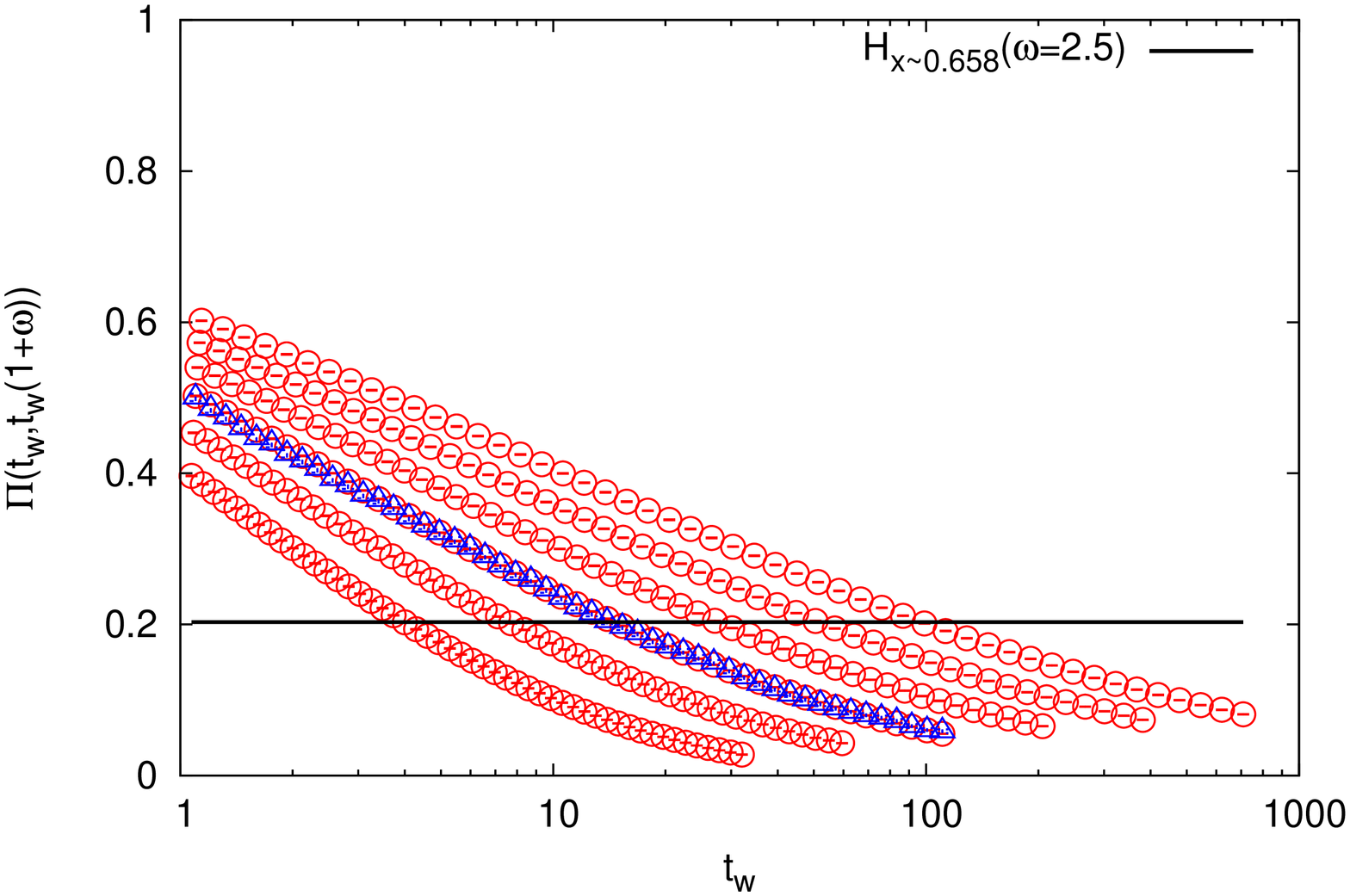}}
 \caption{$\Pi_M[t_w,t_w(1+\omega)]$ with $\omega=0.1$ in \ref{AG_GAU_ADS_T1.58_a0.25_w0.10} and $\omega=2.5$ in \ref{AG_GAU_ADS_T1.58_a0.25_w2.50} for a CTD at $T=1.58/\beta_c$, $a=0.25$, $\rho=0.1$ and $10^6$ trajectories. This set of parameters corresponds to $x\sim 0.658$. Different data series from left to right correspond to system sizes $N=6$, 8, 10, 12, 14 and 16. \label{AG_GAU_ADS_T1.58_a0.25}
 Triangles are for DTD, $N=10$, $10000$ samples, and for large $t_w$ they overlap the $M=1024$ CTD data.}
\end{figure}
We considered $N=6$, 8, 10, 12, 14 and 16, and studied the dynamics on the macroscopic time scale set by $\rho=0.1$. $T=1.58/\beta_c$ and $a=0.25$ which corresponds to $x\sim0.658$, i.e. to the temperature range where aging should be visible: $1/T\sim 0.75$, hence $1/T<\sqrt{2\log 2}/a\sim4.71$ and $1/T>0.49$, which corresponds to Eq. (\ref{bca}) in this case. The averages were performed over $10^6$ samples. The numerical results are shown in Fig. \ref{AG_GAU_ADS_T1.58_a0.25}.
Again we see that it is not possible to clarify the behavior of the system for large $N$ if data are not analyzed as a function of time  in units of $\exp(\gamma N)$ as in Fig. \ref{AG_GAU_ADS_T1.58_a0.25SC}.
\begin{figure}[htb!]
 \centering   
 \subfigure[\label{AG_GAU_ADS_T1.58_a0.25_w0.10SC} ]{\includegraphics[width=0.32\columnwidth, angle=-90]{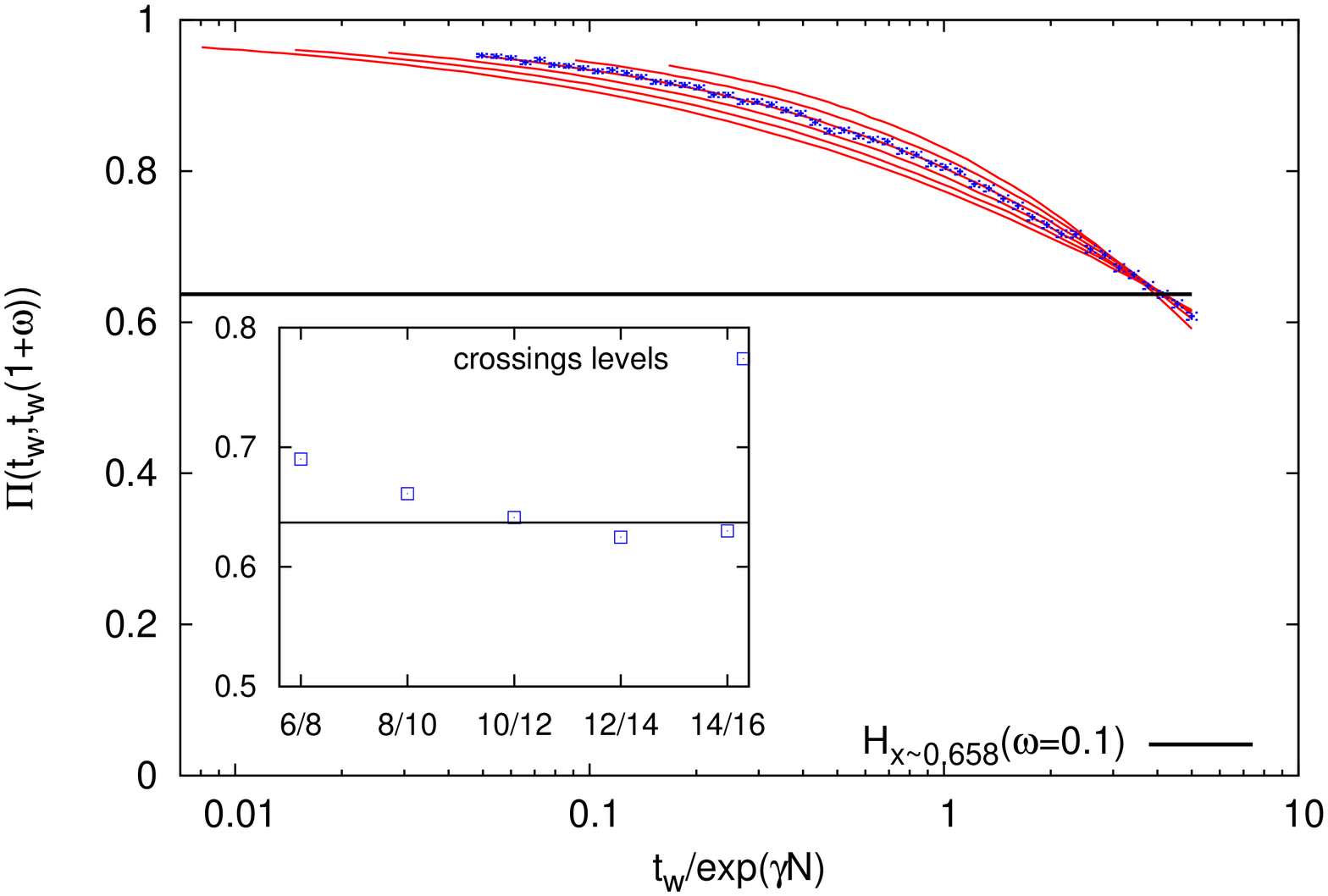}}
 \subfigure[\label{AG_GAU_ADS_T1.58_a0.25_w2.50SC} ]{\includegraphics[width=0.32\columnwidth, angle=-90]{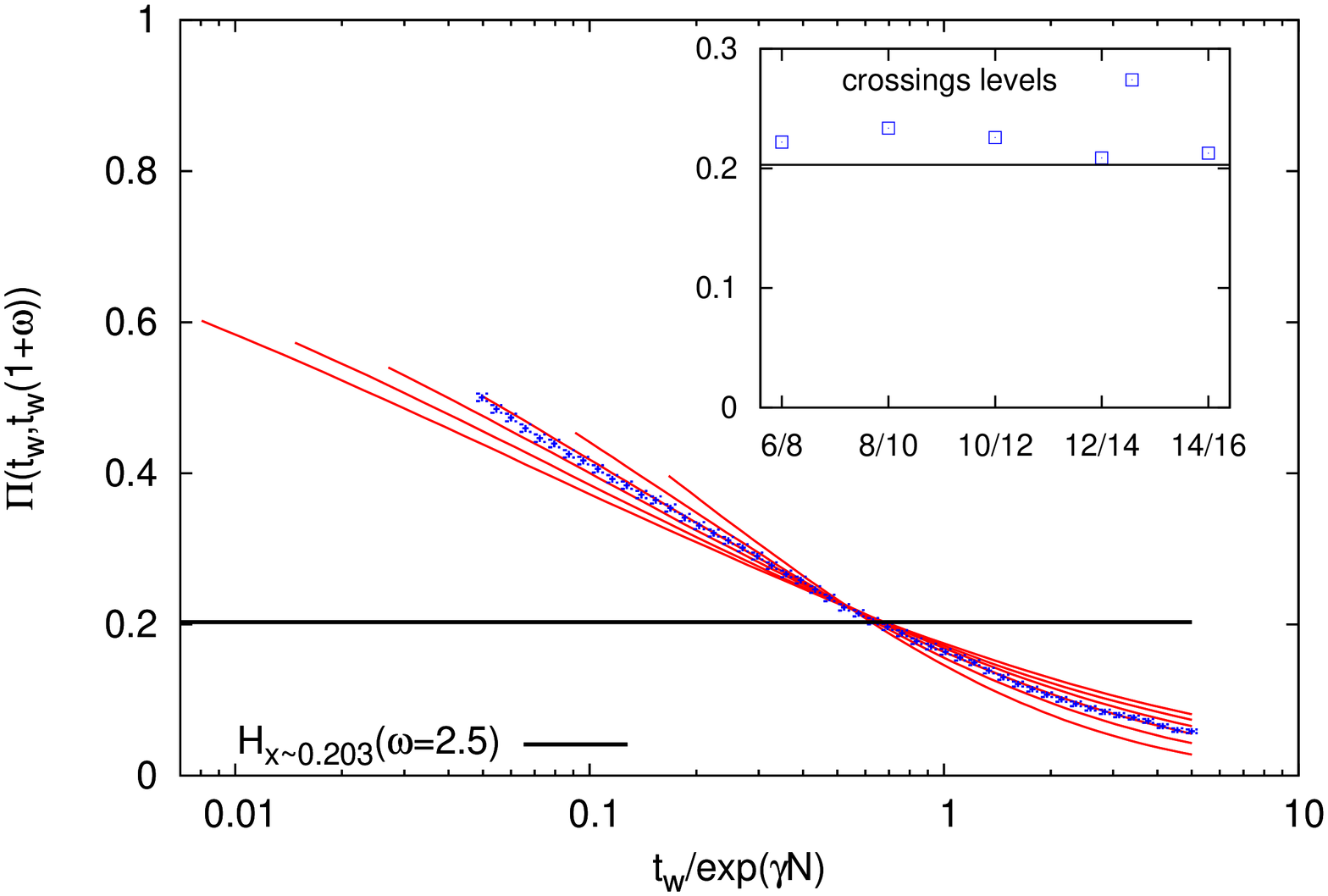}}
 \caption{$\Pi_M[t_w,t_w(1+\omega)]$ from the CTD, also plotted in Fig. \ref{AG_GAU_ADS_T1.58_a0.25}, shown as a function of $t_w/\exp(\gamma N)$. The green dashed line refers to the results obtained with the DTD. The insets show the crossings among consecutive values of $N$, and the convergence to the analytic prediction for the largest values of $N$ is clear. \label{AG_GAU_ADS_T1.58_a0.25SC}}
\end{figure}
After rescaling the time the numerical data clearly show the expected aging result, that is given by the crossing of the curves obtained for different system size. We show in the inset that here finite size corrections are larger than in the $a=0$ case. 

\section{The special case \texorpdfstring{$a=1/2$}{a=1/2} and the Metropolis dynamics}\label{a0.5GTrapSM}
\begin{figure}[htb!]
 \centering   
 \subfigure[\label{AG_EXP_ADS_T0.60_a0.50_w0.50} ]{\includegraphics[width=0.32\columnwidth, angle=-90]{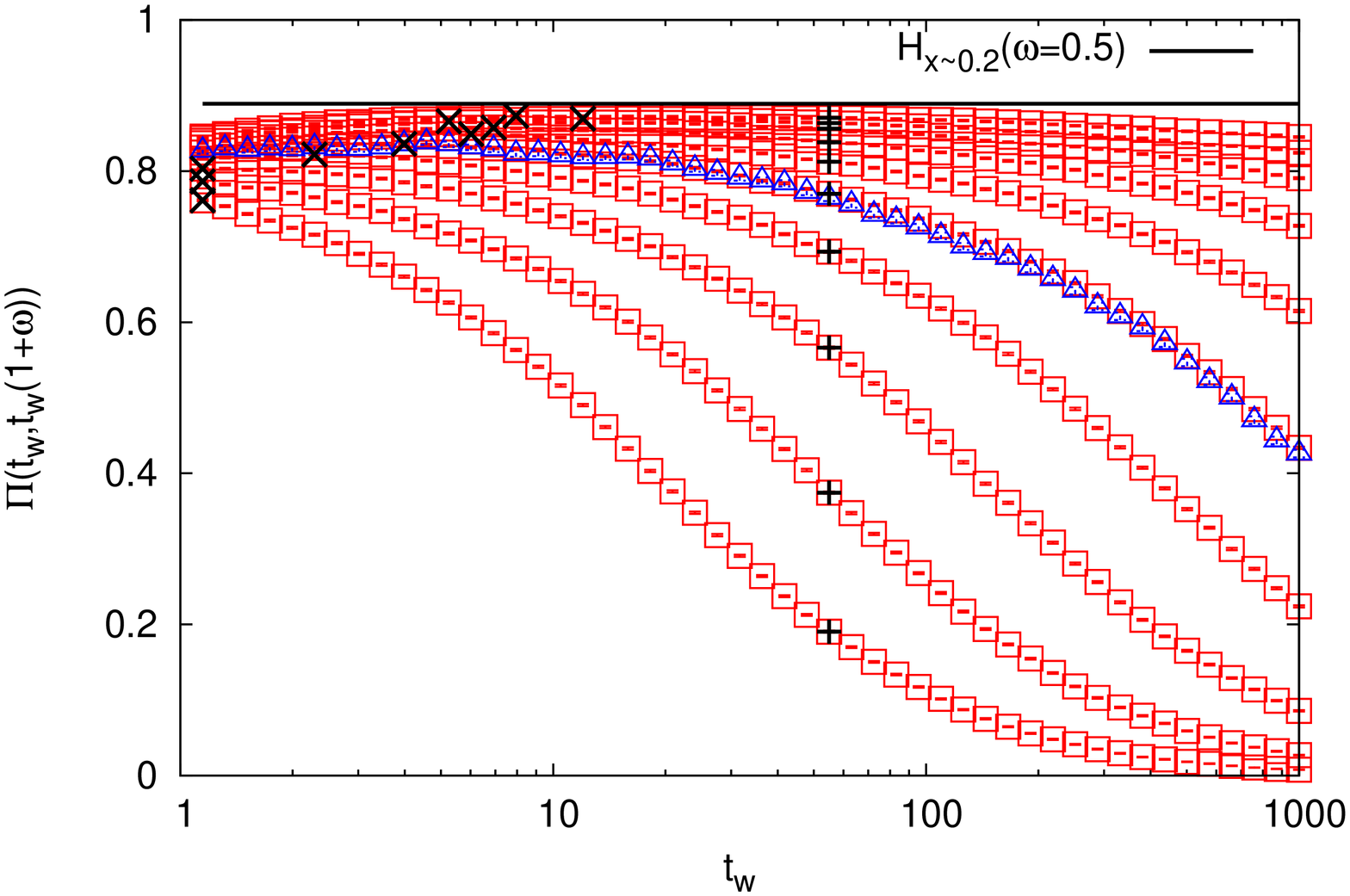}}
 \subfigure[\label{AG_EXP_ADS_T0.60_a0.50_w0.50FSS} ]{\includegraphics[width=0.32\columnwidth, angle=-90]{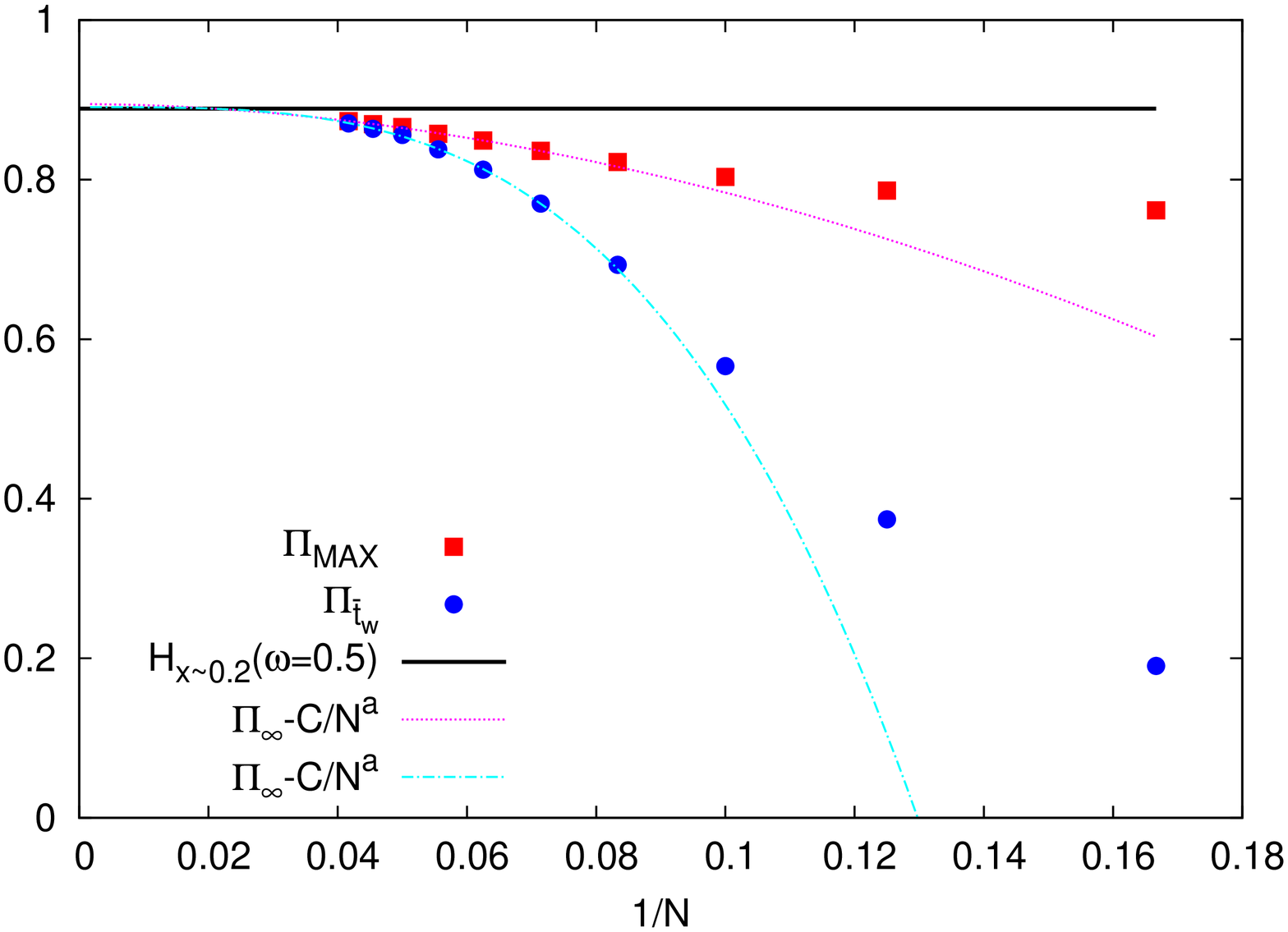}}
 \caption{$\Pi_M[t_w,t_w(1+\omega)]$ with $\omega=0.5$ for a CTD (and DTD, see superimposing triangles) at $T=0.6$, and $a=0.5$ a plotted in \ref{AG_EXP_ADS_T0.60_a0.50_w0.50}. The parameters correspond to $x\sim 0.2$. The crosses superimposed to the data series are $\Pi_{\rm MAX}$ (on the left) and $\Pi_{\rm \overline t_w}$ (on the right). These points are shown in \ref{AG_EXP_ADS_T0.60_a0.50_w0.50FSS} by squares and circles, respectively. The dashed curves are our best fit to $f(N)=\Pi_{\infty}-C/N^a$ for the large system size limit (i.e. $N>12$).}
\end{figure}

\begin{figure}[htb!]
 \centering   
 \includegraphics[width=0.32\columnwidth, angle=-90]{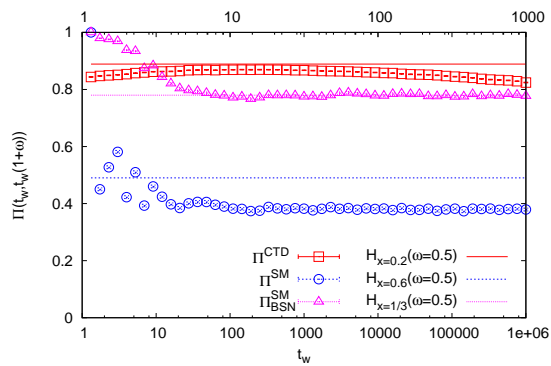}
 \caption{$\Pi[t_w,t_w(1+\omega)]$ and $\Pi_{\rm BSN}[t_w,t_w(1+\omega)]$ for the Step Metropolis at $T=0.6$ with $\omega=0.5$. Different candidates for the trap prediction are shown. The result for the largest system size of a CTD at $T=0.6$, and $a=0.5$ already in \ref{AG_EXP_ADS_T0.60_a0.50_w0.50} is also reported for comparison. These results are also compared with the Arcsin law with parameter $x=(\lambda/\beta-a)/(1-a)$ ($=0.2$ in the considered case), $x=\lambda/\beta=0.6$, $x=2-\beta/\lambda=1/3$.  \label{AG_EXP_NMI_T0.60_w0.50}
}
\end{figure}

In this section we focus on the case $a=1/2$. We have shown that, as expected, the CTD described by Eqs. (\ref{timeaBT}) and (\ref{pija}) and the DTD with transition rates from Eq. (\ref{rija}) produce a statistically similar aging transient, and a clear trap-like aging behavior. The DTD with $E_b=0$ and $a=1/2$ as in a Metropolis dynamics involves symmetrically the initial and final configurations of each step:
\begin{equation}
    r_{i,j}\sim\exp[-\beta\Delta E/2]\;,
\end{equation}
with $\Delta E=E_j-E_i$. The transition rates are the square root of the Metropolis ones with the additional difference that in the Metropolis case if $\Delta E<0$ $r_{i,j}=1$. This could seem an innocent details, but we will see that it dramatically changes the nature of the dynamical process and its aging behavior. Here we are still considering CTD and DTD dynamics that allow the system to move from any configuration to any other one in a single dynamical step. We will also eventually consider a Metropolis dynamics with the same features, i.e. where all configurations are connected from a single elementary move. The Metropolis realization of this fully connected dynamical paradigm was already studied to focus on the aging behavior of the so called Step Model \cite{barmez95,bertin03,cammar15}: here we call it Step Metropolis (SM).\\
We first show the results for our usual CTD and DTD dynamics for the $a=1/2$ case. We show in Fig.~\ref{AG_EXP_ADS_T0.60_a0.50_w0.50} the probability of not changing configuration for a CTD (and in one case for a DTD) with $T=0.6$, $\omega=0.5$ and for Poisson trap models with even $N$ going from 6 to 24, $M=2^9$ and $\lambda=1$. A bigger effort has been done with simulations of larger system sizes because finite size effects are the strongest we have met until now.
For small values of $M$ the numerical results are below the theoretical straight line, but when the system size increases a plateau develops and it progressively reaches higher and higher values. To quantify the large $N$ limit of the numerical results we extract the maxima of the plotted curves, $\Pi_{\rm MAX}(N)$, and the level $\Pi_{\rm \overline t_w}(N)$ that the curves reach at an intermediate fixed time $\overline t_w\sim55$, and we plot them as a function of a system size $N$ (corresponding to the number of configurations $N=\log_2 M$). We fit the data with $N>12$ to $f(N)=\Pi_{\infty}-C/N^a$, and we show the results in Fig.~\ref{AG_EXP_ADS_T0.60_a0.50_w0.50}.  The best fits in the two cases give $\Pi_{\infty}^{\rm MAX}=0.895\pm0.006$ and $\Pi_{\infty}^{\rm \overline t_w}=0.891\pm0.004$, both in remerkable agreement with the theoretical value $0.889$ of the Arcsin law at $\omega=0.5$ with $x= (\lambda/ \beta-a)/ (1-a)=0.2$. Despite a very slow convergence, the trap-like aging behavior predicted in Sec. \ref{predictionsaTrap} is observed for the $a=1/2$ model in the CDT and in the DTD.\\
The study of the SM aging for the Poisson Trap model \cite{cammar15} gives instead very different results, where trap aging at the level of single configurations is absent. In Fig. \ref{AG_EXP_NMI_T0.60_w0.50} we show with blue circles the probability of not changing configurations. The simulations is mimicking the limit of infinite system size because a new configuration, and its energy according to a Poisson distribution, is extracted anew at each step \cite{cammar15} and not taken from a pool of previously extracted energies. This allows not to see a decay from the plateau at large times, making the plateau level clearly visible. \\
$\Pi(t_w,t_w+t)$ is very different from the result of the CDT (red square data points) and does not converge in the large time limit to the Arcsin law with parameter $x=(\lambda/\beta-a)/(1-a)(=0.2\ \rm in\ our\ case)$ nor to the original trap result $x=\lambda/\beta(=0.6\ \rm in\ our\ case)$. Yet, this is not due to finite system's size effect nor to the short times. \\
In order to explore all possibilities let us also note that another candidate for $x$ could be the value that can be extracted from the exponent of the distribution of trapping times \cite{barmez95,bertin03}. In fact for Poisson trap models the exponent of the distribution of trapping times $1+\mu=1+\lambda/\beta$ is straightforwardly reflected into the value of parameter $x$ of the Arcsin law $x=\lambda/\beta$.
However, a SM dynamical rule\footnote{This is also the case for a Glauber DTD with $r_{i,j}\propto \exp(-\beta \Delta E)/(1+\exp(-\beta\Delta E))$.} for $\tau\ll t_w$ gives \cite{barmez95} a $\rho(\tau)$ with exponent $1+\mu=3-\beta/\lambda\neq1+\lambda/\beta$. Hence we should compare the aging results with an Arcsin law with parameter $x=2-\beta/\lambda$. Before doing that and commenting about this comparison let's intuitively explain the origin of such difference in the distribution of  trapping times.\\
The difference in exponents of the trapping time distributions between the trap paradigm and the SM is due to the slightly different nature of the two dynamical processes. In the first case, say in the CTD paradigm, dynamical trapping times always depend on the depth of the currently visited configuration with respect to $E_b$ which acts as an energetic barrier. The dynamics is then always controlled by thermal activated processes. If we think of the DTD paradigm, the transition rates between configurations $i$ and $j$ are all exponentially suppressed compared to the rate of the transition from the configuration with maximum energy $E_{\rm MAX}$ to the configuration with minimum energy $E_{\rm MIN}$. At variance with SM, in the DTD also transition rates towards configurations with lower energies are exponentially suppressed: this turns out to be a crucial difference between the two algorithms. Again formally this means that in the transition rates $r_{i,j}/r_{\rm MAX}$ (with $r_{\rm MAX}$ the transition rate between the configuration with highest energy and the one with lowest energy) the energy difference between $E_i$ and $E_{\rm MAX}$ (besides the one between $E_j$ and $E_{\rm MIN}$) is acting as an energetic barrier for the elementary relaxation processes. Conversely, the SM allows the system to change configuration on a microscopic time scale (transition rate equal to one) as soon as a configuration with smaller energy is found. Note that in the large system size's limit there is always a fraction of neighbouring configurations with smaller energy, but configurations low in energy have less neighbors with even lower energy. 
In this case, the relaxation process becomes slower (leading to large effective trapping times) just because the fraction of low energies neighbours decreases when deep configurations are explored, so it has an entropic origin.  The difference between activated and entropic slowing down becomes clearer in the limit of zero temperature. The out-of-equilibrium dynamics gets completely stuck since the very beginning in the first case (both in the CDT and DTD) and only extremely slow in the long time limit for the SM. Due to this fundamental difference the distribution of trapping times turns out to be controlled by different exponents and in fact the SM was considered to provide an alternative aging paradigm compared to trap models \cite{barmez95}.  \\
Let's now come back to the comparison of the aging results with the Arcsin law predictions. As an apparent confirmation of an irreducible difference between the SM and trap algorithms, also by using $x=2-\beta/\lambda(=1/3\ \rm in\ our\ case)\neq\lambda/\beta$ as a parameter, the Arcsin law does not provide a fair description of numerical data, in principle free from finite size corrections. This would suggest that trap aging paradigm is not at all able to explain the observed numerical results, even when the right exponent of relaxation time's distribution is considered. The rationale for the non-trap nature of the SM can be explained in terms of the absence of the renewal property for this dynamical algorithm. In the SM the choice of new configurations is strongly biased by the configuration the dynamics has reached. In fact the system tends to explore typically configurations with energy slower than the energy of the configuration it has arrived to. This is at variance with the CDT dynamics considered here for which the probability to explore a configuration along the dynamics is independent from the configuration the dynamics has reached. Even in the $a$-generalized case where configuration at lower energies are favoured, this is not dependent on the lastly visited configuration. For this reason the dynamics starts afresh at every single dynamical step. To come back to the SM case, strong correlations arise instead between the energies of subsequent configurations along the dynamics which does not have the renewal feature that a trap dynamics would require. Hence in this case trap model predictions cannot be accurate.\\
However, it has been observed that it exists a range of temperature such that $0.5<\lambda/\beta<1$ where the dynamics explores alternatively deep and shallow configurations effectively giving rise to the existence of dynamical basins \cite{cammar15}. A dynamical basin is defined by the sequence of configurations whose energies lie below a fixed threshold \cite{cammar15}. Every time the dynamical process explores configurations with energy higher than the threshold the system changes basin. In this situation a new way to interpret data can be adopted: the probability of not changing basin $\Pi_{\rm BSN}$ can be defined and it is found to converge to the trap prediction once one sets $x = 2 - \beta / \lambda$, see data for $\Pi_{\rm BSN}$ shown in Fig. \ref{AG_EXP_NMI_T0.60_w0.50}. This result becomes natural if we think that when the system escapes from a basin it has reached shallow configurations at the predefined threshold energy which is independent from how deep the basin was. As such when a basin is abandoned any correlation with the previous history is lost. The renewal property of the dynamics is restored at least on the time scale of basins explorations and the basin description reveals the underlying trap behavior.\\
In conclusion the DTD and a Metropolis dynamics give different results despite the fact that the first was proposed as a generalization of the original trap dynamics to bridge the gap between it and the more common Metropolis algorithm.
Interestingly at first sight both seem not to follow the trap dynamics expectation. \\
Understanding the DTD (and the equivalent CTD) only required a careful finite size scaling study. In particular in the DTD with $a=1/2$ energetic barriers, and activation trapping times, are present for all the target configurations.
Also the dynamical sampling of configurations is not conditioned on the lastly visited one.
Hence the trap aging behaviour must be there, but is hidden by huge finite size corrections.\\
Conversely the SM is strongly influenced by the dynamical paths that do not require jumps towards higher energies. All these paths will have transition rates of single jumps equal to $1$ and strong correlation between configurations subsequently visited. At very low temperatures ($0.5>\lambda/\beta$) transition rates to high energy configurations are always highly suppressed, 
non activated dynamical path always dominate, and trap like dynamics cannot be recovered.
Under certain condition instead, when non activated dynamical path become rare, the transition rate towards higher energies becomes higher than the probability to find them. 
This occurs at intermediate temperatures 
$0.5<\lambda/\beta<1$ where an entropy-energy competition determines the choice of typical dynamical paths. 
In this regime the exploration of the space of configurations is realized through spontaneously formed dynamical basins with effective barriers reaching a common high energy level $E_b$. In this condition the correlations between configurations is still high within the same dynamical basin but configurations belonging to different dynamical basins are totally uncorrelated, hence a trap like aging is finally restored through the description of the dynamics in terms of jumps between different basins.\\
In the two very different dynamics, trap dynamics emerges only after a careful study of the dynamical outcomes. This shows on the one side the broad extension of the trap-like universality class for aging dynamics and on the other side the different motivations for which the trap like aging behavior does not emerge, despite the fact that it is controlling the dynamics.

\section{Discussion and conclusions}
The study of trap dynamics in the case in which transition rates between subsequent configurations depend on both the initial and final configurations aims at bridging the gap between CTD and more common dynamical algorithms. The first ones were used in the original definition of trap models \cite{boucha92,boviha94,boudea94}, but also allowed the first proofs of the extension of the trap paradigm for aging dynamics to a larger number of models \cite{beboga02}. The $a$-generalization of the original CTD shows a simple case where the emergence of trap aging dynamics can be predicted but not always simply numerically tested. The numerical study of this dynamics reveals the tricks that should be adopted to let the trap behavior emerge more clearly without going to prohibitively large system sizes. \\
In the case of the Poisson Trap model we discussed how the long time plateau does not always form at the level where it will be in the large system size.  \\
In the Gaussian Trap model we have shown the importance of knowing the time scale at which the trap-like aging behavior corresponding to a specific parameter of the Arcsin law should emerge. Finite size scaling would not help much without knowing at which time scale the dynamical results coming from different system sizes should be fruitfully compared. \\
Finally the example of the Step Metropolis \cite{cammar15} reviewed here reveals a third mechanism that can hide the emergence of the trap like aging behavior. This mechanism is the arising of dynamical correlations along the dynamics. The interesting result obtained in this case is that even in these situations a well tailored definition of dynamical basins, lumping strongly correlated configurations, could allow the trap predictions to be recovered. \\
It would be interesting to apply these new insights to the numerical study of the REM and possibly $p$-spin models for a better understanding of glassy dynamics in the large time limit where barrier crossing takes place.\\
The REM dynamics has been revealed to be challenging. It refers to a system where energies of the $M$ configurations are i.i.d. random variables distributed according to some meaningful probability distribution (classically Gaussian, but also Poisson) yet configurations ideally represent the available configurations of an array of $N=\log_2M$ spins. As such, the physically meaningful single spin flip dynamics corresponds to a dynamical algorithm with non zero transition rates only for the $N$ transitions that represent a change in one spins of the initial configuration, and null transition rates for the remaining $M-N$ transitions. Since $M\gg N$, this means that the chance of backtracking to already visited configurations increases significantly and introduces correlations along the dynamics.\\
Despite this complication, very recently, it has been shown \cite{cerwas15,gayrar16} that such a dynamics realized by a Metropolis algorithm on a system with Gaussian energies should also show a trap-like aging behavior.  Obtaining numerical evidences of this behavior requires a number of non trivial observations about the dynamical features and an effort to recover the limits (large systems, large observation times, use of basins description) in which dynamical correlations can be neglected \cite{babica17} and trap like behavior emerges numerically. This was successfully achieved for Poisson distribution of the energies \cite{babica17}, but in the Gaussian case the agreement of the numerical results with the trap like behavior is still not clear and it might require a comparison of the dynamics for different system sizes at the right time scale.

\renewcommand{\abstractname}{Acknowledgements}
\begin{abstract}
We thank Marco Baity Jesi, Gerard Ben Arous and Giulio Biroli for interesting discussions. C.C. acknowledges support from the Simons Foundation as affiliate of the collaboration “Cracking the Glass Problem” (No.  454935 to G. Biroli). E.M. has received funding from the European Research Council (ERC) under the European Union’s Horizon 2020 research and innovation programme (grant agreement No [694925]).

\end{abstract}


\end{document}